\documentclass[%
 aip,
  bmf,
 amsmath,amssymb,
 reprint,%
superscriptaddress,nofootinbib,eqsecnum]{revtex4-1}

\usepackage{graphicx}
\usepackage{dcolumn}
\usepackage{bm}

\usepackage[utf8]{inputenc}
\usepackage[T1]{fontenc}
\usepackage{mathptmx}
\usepackage{etoolbox}
\usepackage{hyperref}
\usepackage{xcolor}
\newcommand{\aap}{Astron.\ Astrophys.}

\makeatletter
\def\@email#1#2{%
 \endgroup
 \patchcmd{\titleblock@produce}
  {\frontmatter@RRAPformat}
  {\frontmatter@RRAPformat{\produce@RRAP{*#1\href{mailto:#2}{#2}}}\frontmatter@RRAPformat}
  {}{}
}%
\makeatother
\begin{document}

\preprint{AIP/123-QED}

\title{Optical nonlinearity of cold atomic ensemble\\ driven by strong coherent field in a saturation regime}
\author{A.S. Usoltsev}
 \affiliation{Quantum Technology Centre, M.V.~Lomonosov Moscow State University\\ Leninskiye Gory 1-35, 119991, Moscow, RF}
  \homepage{https://quantum.msu.ru/en}
  \affiliation{Centre for Interdisciplinary Basic Research, HSE University\\ St. Petersburg 190008, RF}
\author{L.V. Gerasimov}
\affiliation{Quantum Technology Centre, M.V.~Lomonosov Moscow State University\\ Leninskiye Gory 1-35, 119991, Moscow, RF}
\affiliation{Centre for Interdisciplinary Basic Research, HSE University\\ St. Petersburg 190008, RF}
\author{A.D. Manukhova}
\affiliation{Department of Optics, Palack\'{y} University, 17 Listopadu 12, 771 46 Olomouc, Czech Republic}
\author{S.P. Kulik}%
\affiliation{Quantum Technology Centre, M.V.~Lomonosov Moscow State University\\ Leninskiye Gory 1-35, 119991, Moscow, RF}%
\author{D.V. Kupriyanov}
 \affiliation{Quantum Technology Centre, M.V.~Lomonosov Moscow State University\\ Leninskiye Gory 1-35, 119991, Moscow, RF}
\affiliation{Centre for Interdisciplinary Basic Research, HSE University\\ St. Petersburg 190008, RF}
\affiliation{Department of Physics, Old Dominion University\\ 4600 Elkhorn Avenue, Norfolk, Virginia 23529, USA}
 \email{kupriyanov@quantum.msu.ru}

\date{\today}

\begin{abstract}
We present a microscopic analysis and evaluation of the dielectric susceptibility of a dielectric medium consisting of vector-type two-energy-level atoms responding on a weak probe mode when the atoms are driven by a strong coherent field. Each atom, in an environment of others, exists as a quasiparticle further structuring a bulk medium. In a limit of dilute atomic gas, the dynamics of each atom follows the Mollow-type nonlinear excitation regime, and the medium susceptibility collectivizes the individual atomic responses to the probe mode. We outline how the collective dynamics can be interpolated up to a dense medium, and we argue from general positions that in such a medium the optical nonlinearity and, in particular, its parametric part could be significantly magnified by manipulating both the coherent pump and the sample density. That indicates certain limitations for potential capabilities of quantum communication protocols utilizing the entangled photons, created by a parametric process, as a main resource of quantum correlations.
\end{abstract}

\maketitle

%

\section{Introduction}

\noindent Optical nonlinear susceptibilities $\chi^{(n)}$, where $n\ge2$, determine how a medium responds to an intense light field by generating new harmonics, mixing frequencies, or creating entangled photon pairs. In particular, in quantum communications, they are responsible for generation of entangled photons in a spontaneous parametric down-conversion or four-wave mixing, for signal transformation in quantum repeaters, for frequency conversion, providing compatibility with telecom networks, and for implementation and rejection of a photon number splitting attack \cite{Lutkenhaus2002}. 

However, the practical implication of optical nonlinearities to quantum communications meets both fundamental and technical limitations: a) Phase noise, accompanying the preparation of photons by a parametric process, reduces the fidelity for a transfer of quantum states. In an example of a quantum repeater, adjusted with a twin-field-quantum-key-distribution protocol, the phase noise from the $\chi^{(2)}$ light source, using LiNbO${}_3$ crystals, limits the communication range by $\sim 500\;km\ $ \cite{Chen2020}; b) At higher pump powers, uncontrolled multi-photon absorption adds a risk of optical damage to any communication protocol; c) In practical realizations, the achievable levels of nonlinearities are limited by available materials and by existing experimental capabilities \cite{Boyd2003,Yu2021,Clinton2024}. 

In this regard, the question naturally arises: what are the physical limits of optical nonlinearities in preparation of multiphoton states feasible for quantum communications?
In general, for unspecified nonlinear optical materials, the answer would depend on many peculiar matter properties and be non-obvious. Nevertheless, to clarify the most critical requirements, one can attempt to find an appropriate theoretical model that allows for a microscopic description under realistic external conditions, and a cold atomic ensemble seems to be a relevant example of such a model.

The fundamental quantum properties of an atom-field interaction in the regime of nonlinear excitation by a strong coherent field were clarified in seminal papers \cite{Rautian1962,Mollow1969} and are now commonly referred to as the Mollow problem.  Decades later, the nonlinear coupling in the four-wave mixing in sodium vapor was utilized in the first experiment, demonstrating the light squeezed by a parametric process, developing under non-degenerate conditions in a cavity, in \cite{Slusher1985} and theoretically described in \cite{Walls1986}. Recently, the concept of using the four-wave mixing in cold alkali-metal atoms as a convenient technical tool for the generation of narrow-band spectrally distributed entangled photons was revived and actively studied in experiments \cite{Shiu2024,Lin2025}. 

Independently, an application of the Mollow problem to light transport, driven by a coherent control field, in ultracold optically dense atomic systems has been investigated experimentally and theoretically in the works \cite{chaneliere2004,Balik2005, Delande2006a,Delande2006b,KSH2017,Binninger2019}. The main attention of those papers was focused on interference phenomena, namely weak and strong localizations of light in disordered atomic gas under conditions of incoherent emission. Specific examples of a light amplifier and a random laser without inversion, driven by a coherent pump on a closed transition in an optically dense atomic gas, were proposed in \cite{Boyd1981,Sorokin2002,Kaiser2009}. Despite these, to our knowledge, the complete vector description of Mollow nonlinearity in an atomic medium is still an open issue for theory.  

Here we aim to construct a nonlinear dielectric susceptibility of a cold atomic ensemble by applying a consistent microscopic approach from the beginning to the end of our derivation sequence, i.e. in a maximally rigorous fashion. Our strategic motivation is to clarify, from fundamental microscopic positions, framed by our model, an option of maximal enhancement of the system nonlinear parametric response on a weak signal mode. For it we consider an example of a four-wave mixing process developing in a dielectric medium consisting of cold atoms having a closed optical transition. For such a system: (i) the light and matter have a strongest coupling, and (ii) the closed transition works maximally effective for creation of nonlinearity with four-wave mixing ranging from weak to saturation regimes.

The paper is organized as follows. In Section \ref{Section_II} and Appendix \ref{Appendix_A} we present a microscopic derivation of the macroscopic Maxwell equations in the Heisenberg formalism, valid in the assumption of long-wavelength dipole gauge. In Section \ref{Section_III} we simplify our consideration by a dilute system and derive the Kubo formula in the case of optical nonlinearity, which expresses the susceptibility tensor via Heisenberg dynamics of the atomic dipoles, driven by a strong coherent field. For such conditions in Section \ref{Section_IV} and Appendix \ref{Appendix_B} we present the consistent derivation of the susceptibility tensor. Our calculations of the Kerr-type and parametric-type nonlinearities as functions of the saturation parameter are presented in Section \ref{Section_V}. An extension of the reported results up to a dense system, structuring a bulk dielectric medium, is outlined in Section \ref{Section_VI}. Finally, we summarize our concluding remarks. 

\section{The macroscopic Maxwell equations\\ in the Heisenberg formalism}\label{Section_II}

\noindent The physical concept, which we shall follow, implies the joint atom-field dynamics of the electromagnetic field existing in an arbitrary state and interacting with an atomic medium. The field can be both strong and weak, so that in certain modes it could saturate the atomic transitions. However, it does not dramatically distort the atomic energy structure and leave it as a stable physical unit. The medium consists of atoms having ground and excited states, belonging to a separated dipole-type optical transition. 

Under these conditions, and for many applications, the interaction process can be approximated by a long-wavelength dipole-type interaction. Since the dipole-type interaction, or dipole gauge, is only approximately valid, its practical use can lead to some variations in the layout of dynamical equations, which are featured by our derivation steps clarified in Appendix \ref{Appendix_A}. That would be most important for dense samples, but even in a dilute regime, when the atoms are separated by distances scaled by their radiation zone, it would be wise to summarize them and clarify our starting position.

The Heisenberg precursor of the macroscopic Maxwell equations can be written in two equivalent forms. Firstly, the Heisenberg equations can be written as coupled operator dynamics for the transverse components of the electric field $\hat{\mathbf{E}}_{\bot}(\mathbf{R},t)$ and magnetic field $\hat{\mathbf{B}}(\mathbf{R},t)$, which are considered as functions of spatial point $\mathbf{R}$ and time $t$
\begin{eqnarray}
\mathrm{rot}\,\hat{\mathbf{B}}(\mathbf{R},t)&=&\frac{1}{c}\,\dot{\hat{\mathbf{E}}}_{\perp}(\mathbf{R},t)+\frac{4\pi}{c}\,\dot{\hat{\mathbf{P}}}_{\perp}(\mathbf{R},t)%
\nonumber\\%
\mathrm{rot}\,\hat{\mathbf{E}}_{\bot}(\mathbf{R},t)&=&-\frac{1}{c}\,\dot{\hat{\mathbf{B}}}(\mathbf{R},t)%
\label{2.1}%
\end{eqnarray}
where 
\begin{equation}
\hat{\mathbf{P}}_{\perp}(\mathbf{R},t)=\sum_{a=1}^{N}\hat{\mathbf{d}}_{\perp}^{(a)}(\mathbf{R},t)%
\label{2.2}
\end{equation}
is transverse component of the microscopic collective dipole density, where the sum is taken over all $N$-atoms of the ensemble. The transversality means orthogonality to the wave vectors $\mathbf{k}$ of each field mode $s=\mathbf{k},\sigma$ with polarization vectors $\mathbf{e}_s\equiv\mathbf{e}_\sigma(\mathbf{k})$ ($\sigma=1,2$), such that in the Schr\"{o}dinger picture for an $a$-th dipole it is given by
\begin{equation}
\hat{\mathbf{d}}_{\perp}^{(a)}(\mathbf{R})=\frac{1}{\cal V}\sum_{s}\mathbf{e}_s\cdot(\mathbf{e}_s\cdot \hat{\mathbf{d}}^{(a)})\,\mathrm{e}^{i\mathbf{k}\cdot(\mathbf{R}-\mathbf{R}_a)}%
\label{2.3}
\end{equation}
where $\mathbf{R}_a$ is the spatial location of the $a$-th atomic dipole. 

Alternatively these equations can be rewritten as
\begin{eqnarray}
\mathrm{rot}\,\hat{\mathbf{B}}(\mathbf{R},t)&=&\frac{1}{c}\,\dot{\hat{\mathbf{D}}}(\mathbf{R},t)%
\nonumber\\%
\mathrm{rot}\,\hat{\mathbf{E}}_{\mathrm{tot}}(\mathbf{R},t)&=&-\frac{1}{c}\,\dot{\hat{\mathbf{B}}}(\mathbf{R},t)%
\label{2.4}%
\end{eqnarray}
where we have introduced the components of the total electric and displacement fields. Their Schr\"{o}dinger originals are respectively given by
\begin{eqnarray}
\hat{\mathbf{E}}_{\mathrm{tot}}(\mathbf{R})&=&\hat{\mathbf{E}}_{\perp}(\mathbf{R})+\sum_{a=1}^{N}\hat{\mathbf{E}}_{\mathrm{dip}}^{(a)}(\mathbf{R})%
\nonumber\\%
\hat{\mathbf{D}}(\mathbf{R})&=&\hat{\mathbf{E}}_{\perp}(\mathbf{R})+ 4\pi\,\hat{\mathbf{P}}_{\perp}(\mathbf{R})%
\nonumber\\%
&=&\hat{\mathbf{E}}_{\mathrm{tot}}(\mathbf{R})+ 4\pi\,\sum_{a=1}^{N}\hat{\mathbf{d}}^{(a)}\,\delta(\mathbf{R}-\mathbf{R}_a)%
\label{2.5}%
\end{eqnarray}
where the longitudinal electric field component are given by (\ref{a.27}). By keeping only the first expectation values of (\ref{2.4}) we arrive at the textbook layout for the macroscopic Maxwell equations of a dielectric medium.

However, in the model considered, only the wave-type equation for the transverse electric field is important and meaningful since (i) the magnetic field has no action on the atoms and (ii) the longitudinal field only copies the dynamics of the dipoles, i.e. their local response on the external field. From (\ref{2.1}) we obtain
\begin{equation}
\triangle \hat{\mathbf{E}}_{\perp}(\mathbf{R},t)-\frac{1}{c^2}\ddot{\hat{\mathbf{E}}}_{\perp}(\mathbf{R},t)=\frac{4\pi}{c^2}\,\ddot{\hat{\mathbf{P}}}_{\perp}(\mathbf{R},t)%
\label{2.6}%
\end{equation}
where the term in the right-hand side defines time derivative of the transverse polarization current.

\begin{figure}[pt]
\includegraphics[width=5cm]{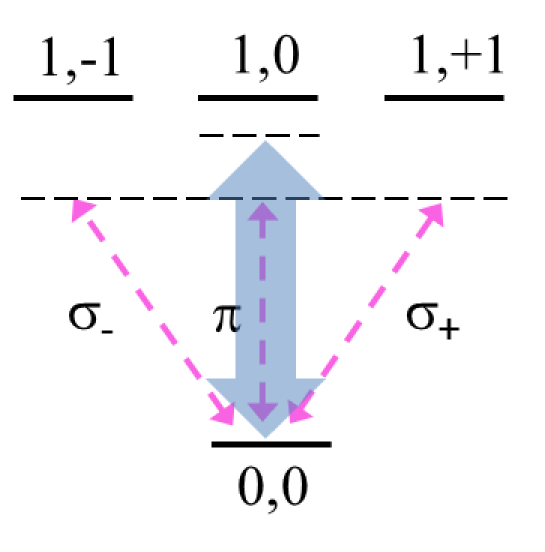}%
\caption{Energy structure of a $V$-type tripod atom constructing the medium used in our simulation of the dielectric sample. The Zeeman states are specified by the angular momentum and its projection. The shaded arrow indicates the strong field quasi-resonant to the reference transition. The three dashed arrows belong to a weak probe impinging on the atom from an arbitrary direction and superposed in the atomic basis. The probe is detuned from both the atomic transition and coherent mode.}
\label{fig1}%
\end{figure}

The Maxwell-Heisenberg equations are obviously incomplete. They should be considered together with complementary equations for the atomic subsystem. In a nonlinear regime that makes the problem extremely hard even for relatively simple energy configurations. In this paper we focus on the Mollow problem with the basic transition diagram shown in Fig.~\ref{fig1}. The medium atom has a single non-degenerate ground state and three excites Zeeman sublevels, i.e. its energy structure is associated with the simplest ${}^{1}S_0\to{}^{1}P_1$ optical transition. It is driven by a strong quasi-resonance coherent field (further referred as the control field), shown by the shaded arrow in the diagram, and, in a steady-state regime, has a nonlinear fluorescence response in three resonance lines known as Mollow triplet. The ensemble of such atoms constructs a disordered atomic medium, where a weak probe light can propagate along an arbitrary direction and with an arbitrary polarization. The latter can be superposed in three orthogonal components defined in the reference frame with its $z$-axis directed along the polarization vector of the strong control mode.   

The suggested Maxwell-Heisenberg approach is generally applicable not only for a dilute gas but also for a highly dense system. 
However, as clarified in Appendix \ref{Appendix_A}, the subtle structure of the system Hamiltonian in its interaction part for proximal dipoles, and physical duality in classification of the field variables inside the medium, namely the difference between electric and displacement components, make the evaluation of the system's response on a weak probe quite challenging when the reference transitions are driven by the control field. 

\section{The Kubo formula}\label{Section_III}

\begin{figure}[pt]
\includegraphics[width=8.5cm]{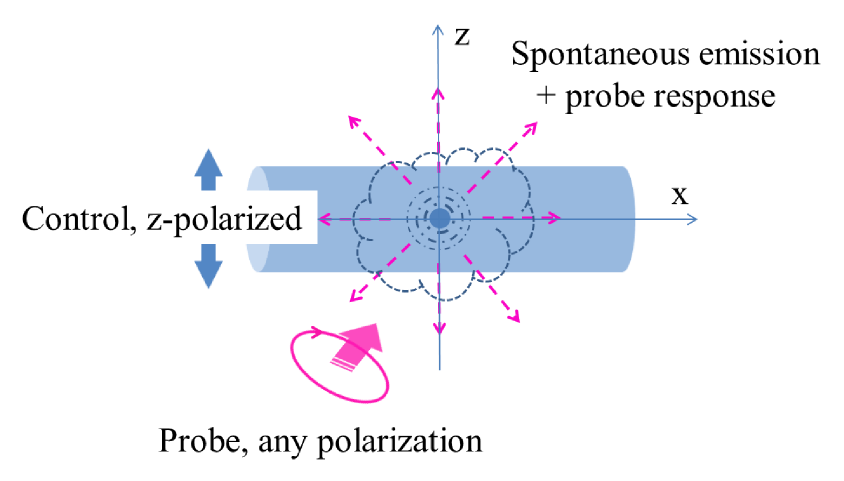}%
\caption{The excitation geometry of an isolated atom, thinkable as an elementary scatterer inside a dilute atomic cloud, and driven by two modes of the strong control and weak probe, as shown in Fig.~\ref{fig1}. The responding field contains both the spontaneous emission and scattered part of the probe mode.}
\label{fig2}%
\end{figure}

\noindent 
The dilute configuration usually allows us to make a crucial simplifying step in resolving the problem.
In the dilute gas, where in the main spatial areas, unoccupied by particles, $\hat{\mathbf{D}}(\mathbf{R})\approx\hat{\mathbf{E}}_{\perp}(\mathbf{R})\approx\hat{\mathbf{E}}_{\mathrm{tot}}(\mathbf{R})$, so that the mutual interference of the proximal dipoles can be safely ignored, we can replace the displacement field by the transverse electric field in the interaction term. For a dilute atomic ensemble, the polarization current responding to the excitation scheme shown in Fig.~\ref{fig1} can be found in a closed form. In Fig.~\ref{fig2} we show the excitation geometry, associated with the transition diagram of Fig.~\ref{fig1}, which we will follow in our derivation.

\subsection{The single atom response}

\noindent 
As is common in the evaluation of a linear response to a weak probe associated with a specific quantized mode, we can
select in the total Hamiltonian the interaction part with it as follows
\begin{equation}
\hat{H}(t)=\hat{H}_0(t)-\hat{\mathbf{d}}\cdot\hat{\pmb{\mathcal E}}%
\label{3.1}%
\end{equation}
where we include in operator $\hat{\pmb{\mathcal E}}$ only one specific probe (which we shall also refer as signal) mode
\begin{equation}
\hat{\pmb{\mathcal E}}(\mathbf{R})=\left(\frac{2\pi\hbar\omega}{{\cal V}}\right)^{1/2}\left[i\mathbf{e}\,a\mathrm{e}^{+i\mathbf{k}\cdot\mathbf{R}}%
-i\mathbf{e}^\ast\,a^{\dagger}\mathrm{e}^{-i\mathbf{k}\cdot\mathbf{R}}\right]%
\label{3.2}%
\end{equation}
where for a sake of notation simplicity and derivation clarity we have omitted any specification of this mode. It is a crucial assumption that elimination of only one mode, defined as $\{\mathbf{e},\omega =ck,\mathbf{k}\}$, from the complete continuum of field modes in (\ref{a.7}) does not affect the original system dynamics driven by the control field and its coupling with the resting part of the field continuum.  The probe field contributes in (\ref{3.1}) at the point of the dipole location $\hat{\pmb{{\mathcal E}}}\equiv\hat{\pmb{{\mathcal E}}}(\mathbf{R}=\mathbf{0})$.

Because of the action of classical control field the internal Hamiltonian $\hat{H}_0(t)$ depends on time even in the Schr\"{o}dinger representation 
\begin{eqnarray}
\hat{H}_0(t)&=&\hat{H}_{\mathrm{Atom}}+\sum_{s}\hbar\omega_s\,(a_s^{\dagger}a_s+1/2)%
\nonumber\\%
&&-\frac{\hbar\Omega_R}{2} |b\rangle\langle a|\mathrm{e}^{-i\omega_c t}- \frac{\hbar\Omega_R}{2} |a\rangle\langle b|\mathrm{e}^{+i\omega_c t}%
-\hat{\mathbf{d}}\cdot\hat{\mathbf{E}}'%
\nonumber\\%
\label{3.3}%
\end{eqnarray}
where $\omega_c$ is the frequency of the coherent control field, and we have implied the the rotating wave approximation (RWA) for interaction with all the field modes distributed near the reference atomic optical transition with frequency $\omega_{b}\equiv\omega_0$, see definition of the atomic Hamiltonian (\ref{a.2}) (with its self-energy part neglected). Here $\hbar\Omega_R=d_0E_0$ denotes the Rabi frequency $\Omega_R$ of the control field for coupling the states $|a\rangle=|0,0\rangle$ and $|b\rangle=|1,0\rangle$, see Fig.~\ref{fig1}, and $d_0=d_{ab}=d_{ba}>0$ is the matrix element of the transition dipole moment, which without loss of generality is assumed to be real and positive. Because of free choice in definition of the initial time, we can assume $\Omega_R>0$, i.e., fix the complex amplitude $E_0$ as real and positive as well. In the last term, responding for interaction with the quantized and initially unoccupied vacuum modes, we have eliminated the interaction with the control and probe modes.  For clarity, the absence of these modes in the expansion (\ref{a.7}) is superscribed by the prime sign in the last term of (\ref{3.3}) with $\hat{\mathbf{E}}\equiv\hat{\mathbf{E}}(\mathbf{R}=\mathbf{0})$.

Define the evolutionary operator of the Hamiltonian  (\ref{3.3}) 
\begin{equation}
U(t,0)=T\exp\left[-\frac{i}{\hbar}\int_0^{t}\hat{H}_0(t')\,dt'\right]%
\label{3.4}%
\end{equation}
where $T$ is the chronologically ordering operator acting on each operator product appearing from the exponent expansion: $\hat{H}_0(t_1)\hat{H}_0(t_2)\ldots\hat{H}_0(t_n)$, such that $t_1>t_2>\ldots >t_n$. Then we can define the interaction representation for any system operator. In particular, the dipole moment transforms as follows
\begin{equation}
\hat{\mathbf{d}}(t)=U^{\dagger}(t,0)\,\hat{\mathbf{d}}\,U(t,0)%
\label{3.5}%
\end{equation}
In accordance with our above assumptions, we neglect those part of the emitted nonlinear fluorescence, which would overlap the probe mode, and accept
\begin{eqnarray}
\hat{\pmb{\mathcal E}}(\mathbf{R},t)&=&U^{\dagger}(t,0)\,\hat{\pmb{{\mathcal E}}}(\mathbf{R})\,U(t,0)%
\nonumber\\%
&=&\left(\frac{2\pi\hbar\omega}{{\cal V}}\right)^{1/2}\left[i\mathbf{e}\,a(t)\,\mathrm{e}^{+i\mathbf{k}\cdot\mathbf{R}}%
-i\mathbf{e}^\ast\,a^{\dagger}(t)\,\mathrm{e}^{-i\mathbf{k}\cdot\mathbf{R}}\right]%
\nonumber\\%
&\equiv&\left(\frac{2\pi\hbar\omega}{{\cal V}}\right)^{1/2}\left[i\mathbf{e}\,a\,\mathrm{e}^{-i\omega t+i\mathbf{k}\cdot\mathbf{R}}%
-i\mathbf{e}^\ast\,a^{\dagger}\,\mathrm{e}^{+i\omega t-i\mathbf{k}\cdot\mathbf{R}}\right]%
\nonumber\\%
\label{3.6}%
\end{eqnarray}
These basic transformations allow us to express the system dynamics in the interaction representation.

Indeed the basic Schr\"{o}dinger equation for an arbitrary system state $|t\rangle_{S}$ in the Schr\"{o}dinger representation reads
\begin{equation}
i\hbar\frac{\partial}{\partial t}|{t}\rangle_{S}=\left[\hat{H}_0(t)-\hat{\mathbf{d}}\cdot\hat{\pmb{\cal E}}\right]|t\rangle_{S}%
\label{3.7}%
\end{equation}
With making use of (\ref{3.4}) we can link it with interaction representation 
\begin{equation}
|{t}\rangle_{S}=U(t,0)|{t}\rangle_{I}%
\label{3.8}%
\end{equation}
and construct the Schr\"{o}dinger equation in the interaction representation
\begin{equation}
i\hbar\frac{\partial}{\partial t}|{t}\rangle_{I}=-\hat{\mathbf{d}}(t)\cdot\hat{\pmb{\mathcal E}}(t)|t\rangle_{I}%
\label{3.9}%
\end{equation}
where $\hat{\mathbf{d}}(t)$ and $\hat{\pmb{\mathcal E}}(t)$ are respectively given by (\ref{3.5}) and (\ref{3.6}). 

We can completely eliminate the state dependence on time by introducing the evolutionary transformation from the interaction to Heisenberg picture
\begin{equation}
S(t,0)=T\exp\left[+\frac{i}{\hbar}\int_0^{t}\hat{\mathbf{d}}(t')\cdot\hat{\pmb{\mathcal E}}(t')\,dt'\right]%
\label{3.10}%
\end{equation}
and
\begin{equation}
|{t}\rangle_{I}=S(t,0)|\rangle_{H}%
\label{3.11}%
\end{equation}
constitutes the time-independent state $|\rangle_{H}$ in the Heisenberg description of the system dynamics.

Eventually the dipole operator in the Heisenberg representation is given by
\begin{eqnarray}
\hat{\mathbf{d}}_{H}(t)&=&S^{\dagger}(t,0)\,\hat{\mathbf{d}}(t)\,S(t,0)%
\nonumber\\%
&\approx&\hat{\mathbf{d}}(t)-\frac{i}{\hbar}\int_0^t\left[\hat{\mathbf{d}}(t')\cdot\hat{\pmb{\mathcal E}}(t'),\,\hat{\mathbf{d}}(t)\right]\,dt'+\ldots
\nonumber\\%
\label{3.12}%
\end{eqnarray}
 where in the second line we have implied the first order approximation in expansion of the evolutionary operator. For a sake of notation convenience here we additionally subscribe this operator by "$H$" and reserve lighter notation $\hat{\mathbf{d}}(t)$ for the interaction dynamics, which is actually important for us and will be further tracked throughout our derivation.

 The Kubo formula follows directly from (\ref{3.12}), written in terms of expectation values, and assumes a given dipole position. Throughout our consideration, we treat the position of atom[s] classically. Thus, in order to link it with the dipole density, we have to multiply (\ref{3.12}) by $\delta(\mathbf{R})$. The mean dipole density (local polarization) of a single atom is given by
 \begin{equation}
 \mathbf{P}(\mathbf{R},t)=\langle\hat{\mathbf{d}}_{H}(t)\rangle\,\delta(\mathbf{R})\equiv\bar{\mathbf{d}}_{H}(t)\,\delta(\mathbf{R})%
 \label{3.13}%
 \end{equation}
and the mean probe field
\begin{equation}
\pmb{\mathcal E}(\mathbf{R},t)=\langle\hat{\pmb{\mathcal E}}(\mathbf{R},t)\rangle%
 \label{3.14}%
 \end{equation}
Then (\ref{3.12}) leads to
\begin{equation}
P_{i}(\mathbf{R},t)=\bar{d}_{i}(t)\,\delta(\mathbf{R})+\int_{-\infty}^{t}\alpha_{ij}(\mathbf{R};t,t')\,{\cal E}_{j}(\mathbf{R},t')\,dt'%
 \label{3.15}%
 \end{equation}
where
\begin{equation}
\alpha_{ij}(\mathbf{R};t,t')=\frac{i}{\hbar}\left\langle\left[\hat{d}_{i}(t),\hat{d}_{j}(t')\right]\right\rangle\,\delta(\mathbf{R})%
\label{3.16}%
\end{equation}
defines the polarizability tensor for a single atom in the Cartesian basis with $i,j=x,y,z$ and with default sum in (\ref{3.15}) over the repeated tensor index. 

The relations (\ref{3.15}) and (\ref{3.16}) derived for a linear response of the dipole polarization on a probe mode constitute the Kubo formula in the case of a single atom, see \cite{LandauLifshitz-V}. However, the presence of the first term in (\ref{3.15}) is a consequence of nonlinear dynamics of the atomic dipole driven by the Hamiltonian (\ref{3.3}). In the steady state excitation condition, formally approached by $t\to\infty$, and in assumption that $\alpha_{ij}(\mathbf{R};t,t')\to 0$ if $t-t'\to\infty$, the integral in (\ref{3.15}) becomes independent of its lower time argument, formally extended to $-\infty$.


\subsection{Generalization on a multi-particle system}

 \noindent Instead of (\ref{3.1}) we have
 \begin{equation}
\hat{H}(t)=\hat{H}_0(t)-\sum_{a=1}^{N}\hat{\mathbf{d}}^{(a)}\cdot\hat{\pmb{\mathcal E}}_{a}%
\label{3.17}%
\end{equation}
where $\hat{\pmb{\mathcal E}}_{a}$ is given by (\ref{3.2}) at the point of $a$-th atom location: $\hat{\pmb{\mathcal E}}_{a}\equiv\hat{\pmb{\mathcal E}}(\mathbf{R}=\mathbf{R}_a)$. The undisturbed Hamiltonian is given by
\begin{eqnarray}
\hat{H}_0(t)&=&\sum_{a=1}^{N}\hat{H}_{\mathrm{Atom}}^{(a)}+\sum_{s}\hbar\omega_s\,(a_s^{\dagger}a_s+1/2)%
\nonumber\\%
&-&\sum_{a=1}^{N}\left[\frac{\hbar\Omega_R}{2}\,|b\rangle\langle a|^{(a)}\,\mathrm{e}^{-i\omega_c t+i\mathbf{k}_c\cdot\mathbf{R}_a}\right.%
\nonumber\\%
&&\left.+\frac{\hbar\Omega_R}{2}\,|a\rangle\langle b|^{(a)}\,\mathrm{e}^{+i\omega_c t-i\mathbf{k}_c\cdot\mathbf{R}_a}\right]%
\nonumber\\%
&&-\sum_{a=1}^{N}\hat{\mathbf{d}}^{(a)}\cdot\hat{\mathbf{E}}'(\mathbf{R}_a)%
\label{3.18}%
\end{eqnarray}
where here and throughout we superscribe by the atom's number its internal Hamiltonian and dyad-type operators. 
The critical aspect in equation (\ref{3.18}) is the spatial dependence of the interaction terms for each particular atomic dipole.

The Rabi frequencies are changed along the sample, so that $\Omega_R=\Omega_R(\mathbf{R}_a)$. However, in dilute systems the scattering and depletion of the control field is a weak process. That means that for sufficiently long distances inside the sample it can be assumed to have a uniform profile. Its spatial dependence along the sample can be approximately taken into account by spatial parametrization of the sample susceptibility in the finally derived equations.  

Other transformations are straightforward. The evolutionary operator $U(t,0)$ is defined by (\ref{3.4}) and transformation (\ref{3.5}) should be specified for each dipole
\begin{equation}
\hat{\mathbf{d}}^{(a)}(t)=U^{\dagger}(t,0)\,\hat{\mathbf{d}}^{(a)}\,U(t,0)%
\label{3.19}%
\end{equation}
and (\ref{3.6}) is unchanged. Equations (\ref{3.7})-(\ref{3.11}) should be generalized for the collection of atoms and $|t\rangle_{S}$, $|t\rangle_{I}$, and $|\rangle_{H}$ would describe the quantum states of the global system. Evolutionary operator (\ref{3.10}), responsible for interaction of the dipoles with the probe mode, generalizes as
\begin{equation}
S(t,0)=T\exp\left[+\frac{i}{\hbar}\int_0^{t}\sum_{b=1}^{N}\hat{\mathbf{d}}^{(b)}(t')\cdot\hat{\pmb{\mathcal E}}_{b}(t')\,dt'\right]%
\label{3.20}%
\end{equation}
and instead of (\ref{3.12}) we arrive at
\begin{eqnarray}
\hat{\mathbf{d}}_{H}^{(a)}(t)&=&S^{\dagger}(t,0)\,\hat{\mathbf{d}}^{(a)}(t)\,S(t,0)%
\nonumber\\%
&\approx&\hat{\mathbf{d}}^{(a)}(t)-\frac{i}{\hbar}\sum_{b=1}^{N}\int_0^t\left[\hat{\mathbf{d}}^{(b)}(t')\cdot\hat{\pmb{\mathcal E}}_{b}(t'),\,\hat{\mathbf{d}}^{(a)}(t)\right]\,dt'%
\nonumber\\%
&&+\ldots
\label{3.21}%
\end{eqnarray}
The operators of atomic dipoles, dressed by interaction with the control field, are delocalized and sharing all the positions $a,b=1,2\ldots N$. The interaction of atoms with light, emitted as nonlinear fluorescence, can entangle them. In other words, the operators $\hat{\mathbf{d}}^{(b)}(t')$ and $\hat{\mathbf{d}}^{(a)}(t)$ for $a\neq b$ are generally uncommuted and correlated.

Eq. (\ref{3.13}) generalizes to
\begin{equation}
 \mathbf{P}(\mathbf{R},t)=\sum_{a=1}^{N}\langle\hat{\mathbf{d}}_{H}^{(a)}(t)\rangle\,\delta(\mathbf{R}-\mathbf{R}_a)\equiv\sum_{a=1}^{N}\bar{\mathbf{d}}_{H}^{(a)}(t)\,\delta(\mathbf{R}-\mathbf{R}_a)%
 \label{3.22}%
 \end{equation}
and Eq. (\ref{3.21}) leads to
\begin{eqnarray}
P_{i}(\mathbf{R},t)&=&\sum_{a=1}^{N}\bar{d}_{i}^{(a)}(t)\,\delta(\mathbf{R}-\mathbf{R}_a)%
\nonumber\\%
&&+\int_{-\infty}^{t}dt'\int d^3R'\,\alpha_{ij}(\mathbf{R},t;\mathbf{R}',t')\,{\mathcal E}_{j}(\mathbf{R}',t')%
\nonumber\\%
\label{3.23}%
 \end{eqnarray}
where
\begin{eqnarray}
\lefteqn{\alpha_{ij}(\mathbf{R},t;\mathbf{R}',t')=\sum_{a,b}\alpha_{ij}^{(a,b)}(\mathbf{R},t;\mathbf{R}',t')}%
\nonumber\\%
&&\equiv\frac{i}{\hbar}\sum_{a,b}\left\langle\left[\hat{d}_{i}^{(a)}(t),\hat{d}_{j}^{(b)}(t')\right]\right\rangle\,%
\delta(\mathbf{R}-\mathbf{R}_a)\,\delta(\mathbf{R}'-\mathbf{R}_b)%
\nonumber\\%
\label{3.24}%
\end{eqnarray}
which shows, in general, a non-local response of the medium polarization on the driving weak probe, similar but not identical to the effect of spatial dispersion. The spatial correlations are created by nonlinear dynamics of the dipoles and can exist only for a particular and frozen configuration of atoms.


\subsection{Thermal averaging}

\noindent In reality, the positions of the atomic dipoles are random.
Even if we consider an ensemble of cold atoms,
so that the Doppler shifts are negligible, their spatial configuration is permanently changing. Then reasonable to assume that any interatomic correlations can be canceled out and the dilute system celebrates the single atom response, expressed by (\ref{3.15}) and (\ref{3.16}):
\begin{equation}
P_{i}(\mathbf{R},t)=\bar{d}_{i}^{(\mathbf{R})}(t)\,n(\mathbf{R})+\int_{-\infty}^{t}\chi_{ij}(\mathbf{R};t,t')\,{\cal E}_{j}(\mathbf{R},t')\,dt'%
 \label{3.25}%
 \end{equation}
where
\begin{equation}
\chi_{ij}(\mathbf{R};t,t')=\frac{i}{\hbar}\left\langle\left[\hat{d}_{i}^{(\mathbf{R})}(t),\hat{d}_{j}^{(\mathbf{R})}(t')\right]\right\rangle\,n(\mathbf{R})%
\label{3.26}%
\end{equation}
is the medium dielectric susceptibility, if reduced per single atom, would be the same as a dipole polarizability $\alpha_{ij}(\mathbf{R};t,t')$.
Here, $n=n(\mathbf{R})$ is a smoothed density distribution, expressed by a given uniform spatial profile, and the dipole $\hat{\mathbf{d}}^{(\mathbf{R})}(t)$ is driven by Hamiltonian (\ref{3.18}) and is considered a function of the smoothed spatial coordinate $\mathbf{R}_a\to\mathbf{R}$.

\section{The sample susceptibility vs. Mollow problem}\label{Section_IV}

\noindent The derived Kubo formula gives us the mathematical resource needed to evaluate the sample susceptibility. However, this can be done only within certain assumptions which validity is provided by our model and clarified below.

\subsection{Basic definitions and transformations}

\noindent For an arbitrary atomic ensemble the Hamiltonian (\ref{3.18}) cannot be split into the sum of partial terms (\ref{3.3})
\begin{equation}
\hat{H}_0(t)\neq\sum_{a=1}^{N}\left.\hat{H}_0^{(\mathbf{R})}(t)\right|_{\mathbf{R}\to\mathbf{R}_a}%
\label{4.1}
\end{equation}
because the field Hamiltonian is globally unique and the quantized field exists as one undivided environment for all the atoms. The separation of proximal dipoles physically limits a scale for the quantization box in (\ref{a.7}), where it is assumed to be infinitely large.

For the term of interaction with the quantized field in (\ref{3.18}), in assumption of the RWA approach, we can define
\begin{eqnarray}
|b\rangle\langle b|&=&|b\rangle\langle b|^{(\mathbf{R})}%
\nonumber\\%
|b\rangle\langle a|&=&|b\rangle\langle a|^{(\mathbf{R})}\,\mathrm{e}^{+i\mathbf{k}_c\cdot\mathbf{R}}%
\nonumber\\%
|a\rangle\langle b|&=&|a\rangle\langle b|^{(\mathbf{R})}\,\mathrm{e}^{-i\mathbf{k}_c\cdot\mathbf{R}}%
\label{4.2}%
\end{eqnarray}
which shifts the frame origin to the atom's location.

We have arrived at inconvenient observation: the operator transformations (\ref{4.2}), aiming to boost the frame origin to a particular atom of the ensemble, map the dependence on its location on the field operators (\ref{a.7}). For a single atom it would not be a problem, since the extra phase shifts induced in the field operators can be incorporated to a fundamental phase uncertainty in the definition of the canonic operators $a_s$ and $a_s^{\dagger}$. However, in a system consisting of many particles, the field modes should be specified as a unique set identical to either atom. We clearly see that the light-matter interaction is an intrinsically cooperative and configuration-dependent process. Hopefully, for a dilute atomic ensemble, the problem can be softened and resolved by the following arguments.

Imagine the atomic ensemble as loaded into a three-dimensional grid built up by physically scaled quantization boxes -- $L\sim n_0^{-1/3}$ -- where $n_0$ is the density of atoms. Thus, on average, there is only one atom per each quantization box. The discretization of $\mathbf{k}$ by the quantization modes is now given by a "coarse-grained" scaling in the momentum space. The exponents in (\ref{a.7}) become periodic functions, which argument gets maximal increment $k_xL+k_yL+k_zL$ $\mod(2\pi)$ within each box. In addition, we can fit the length $L$ in such a way that the control field would also fulfill these revised quantization conditions.  Then the meaningful variation of the exponent arguments in the interaction term of (\ref{3.18}), estimated as $\sim\Delta\lambda/\lambda\,O(2\pi)\ll 1$, where $\Delta\lambda=\lambda-\lambda_c$ is a deviation of the wavelength justified by RWA, can be neglected. 

We conclude that for any atom, taken from the ensemble, its dynamics is driven by the Hamiltonian (\ref{3.18}) where the atom's location can be shifted to the frame origin with saving the uniform definition of the basic parameters such as Rabi frequency, transition dipole moments, etc. As a consequence, the Schr\"{o}dinger precursors of its dyad operators (\ref{4.2}) can evolve up to their Heisenberg dynamics being considered at the frame origin. Then, in the principal frame, with axes $x,y,z$, see Fig.~\ref{fig2}, the commutator of the dipole operators in (\ref{3.26}) can be expanded in terms of their frequency components as follows
\begin{equation}
\left[\hat{d}_{x}^{(\mathbf{R})}(t),\hat{d}_{x}^{(\mathbf{R})}(t')\right]=\left[\hat{d}_{x}^{(+)}(t),\hat{d}_{x}^{(-)}(t')\right]%
+\left[\hat{d}_{x}^{(-)}(t),\hat{d}_{x}^{(+)}(t')\right]%
\label{4.3}%
\end{equation}
and
\begin{equation}
\left[\hat{d}_{y}^{(\mathbf{R})}(t),\hat{d}_{y}^{(\mathbf{R})}(t')\right]=\left[\hat{d}_{y}^{(+)}(t),\hat{d}_{y}^{(-)}(t')\right]%
+\left[\hat{d}_{y}^{(-)}(t),\hat{d}_{y}^{(+)}(t')\right]%
\label{4.4}%
\end{equation}
and
\begin{eqnarray}
\lefteqn{\left[\hat{d}_{z}^{(\mathbf{R})}(t),\hat{d}_{z}^{(\mathbf{R})}(t')\right]=\left[\hat{d}_{z}^{(+)}(t),\hat{d}_{z}^{(-)}(t')\right]%
+\left[\hat{d}_{z}^{(-)}(t),\hat{d}_{z}^{(+)}(t')\right]}%
\nonumber\\%
&&+\left[\hat{d}_{z}^{(+)}(t),\hat{d}_{z}^{(+)}(t')\right]\,\mathrm{e}^{+2i\mathbf{k}_c\cdot\mathbf{R}}%
+\left[\hat{d}_{z}^{(-)}(t),\hat{d}_{z}^{(-)}(t')\right]\,\mathrm{e}^{-2i\mathbf{k}_c\cdot\mathbf{R}}%
\nonumber\\%
\label{4.5}%
\end{eqnarray}
Here we have taken into account that the direct action of the control field does not concern the upper states $|1,\pm 1\rangle$, and $\hat{d}_{x,y}(t)$ obey the dynamics, which is only indirectly affected by the coupling of the control field on $|a\rangle\to|b\rangle$ transition. Hence, the identical frequency components of $\hat{d}_{x}(t)=\hat{d}_{x}^{(+)}(t)+\hat{d}_{x}^{(-)}(t)$ and $\hat{d}_{y}(t)=\hat{d}_{y}^{(+)}(t)+\hat{d}_{y}^{(-)}(t)$ commute at any time.

Express the dipole operators by the dyad operators. On the active transition we have
\begin{eqnarray}
\hat{d}_{z}^{(+)}(t)&=&d_0\,|a\rangle\langle b|(t)%
\nonumber\\%
\hat{d}_{z}^{(-)}(t)&=&d_0\,|b\rangle\langle a|(t)%
\label{4.6}%
\end{eqnarray}
where the dynamics of the operators in the right-hand side can be found from the solution of the Mollow problem within its two-level approximation. Next define the following two undisturbed upper states
\begin{eqnarray}
|x\rangle&\equiv&\frac{1}{\sqrt{2}}\left[|1,-1\rangle-|1,+1\rangle\right]\to\frac{1}{2}\sqrt{\frac{3}{\pi}}\sin\theta\,\cos\phi%
\nonumber\\%
|y\rangle&\equiv&\frac{i}{\sqrt{2}}\left[|1,-1\rangle+|1,+1\rangle\right]\to\frac{1}{2}\sqrt{\frac{3}{\pi}}\sin\theta\,\sin\phi%
\nonumber\\%
\label{4.7}%
\end{eqnarray}
which, together with the state
\begin{equation}
|b\rangle\equiv|z\rangle\to\frac{1}{2}\sqrt{\frac{3}{\pi}}\cos\theta%
\label{4.8}%
\end{equation}
span the three dimensional upper level subspace onto a vector extension of the Mollow problem. Here $\theta,\phi$ is the spherical angle, associated with the frame of Fig.~\ref{fig2}, and $d_{ab}\equiv d_{az}=d_{ax}=d_{ay}=d_0>0$. Then similarly to (\ref{4.6}) we define
\begin{eqnarray}
\hat{d}_{x}^{(+)}(t)&=&d_0\,|a\rangle\langle x|(t)%
\nonumber\\%
\hat{d}_{x}^{(-)}(t)&=&d_0\,|x\rangle\langle a|(t)%
\label{4.9}%
\end{eqnarray}
and
\begin{eqnarray}
\hat{d}_{y}^{(+)}(t)&=&d_0\,|a\rangle\langle y|(t)%
\nonumber\\%
\hat{d}_{y}^{(-)}(t)&=&d_0\,|y\rangle\langle a|(t)%
\label{4.10}%
\end{eqnarray}
The operators in the right-hand sides of (\ref{4.9}) and (\ref{4.10}) can be found via solution of the extended Mollow problem, which we discuss in the next section.

\subsection{Three dimensional extension of the Mollow problem}

\noindent We follow and generalize the calculation approach of \cite{CohTann92}. Let us define the following, slow varying in time, operators for the atom
\begin{eqnarray}
\hat{\sigma}_{+}(t)&=&|b\rangle\langle a|(t)\,\mathrm{e}^{-i\omega_ct}
\nonumber\\%
\hat{\sigma}_{-}(t)&=&|a\rangle\langle b|(t)\,\mathrm{e}^{+i\omega_ct}
\nonumber\\%
\hat{\sigma}_{Z}(t)&=&\frac{1}{2}\big[|b\rangle\langle b|(t) - |a\rangle\langle a|(t)\big]%
\label{4.11}%
\end{eqnarray}
where the lower sign indices ${}_\pm$, associated with either creation or annihilation event, should not be confused with the upper indices ${}^{(\pm)}$, associated with the operator's frequency components. and the capital $Z$-index, specifying the atomic pseudospin, should not be confused with the vector component $z$.

Then for the main excitation channel $|a\rangle\to|b\rangle$ the Heisenberg-Langevin dynamical equations for the atomic operators subsequently read
\begin{eqnarray}
{\dot {\hat \sigma}}_+(t)&=&\left(-{\rm i} \Delta - \frac{\gamma}{2}\right){\hat \sigma}_+(t) + {\rm i} \Omega_R\,{\hat \sigma}_{\mathrm{Z}}(t) +  {\hat F}_+(t)%
\nonumber\\%
{\dot {\hat \sigma}}_-(t)&=&\left(+{\rm i} \Delta - \frac{\gamma}{2}\right){\hat \sigma}_-(t) - {\rm i} \Omega_R\,{\hat \sigma}_{\mathrm{Z}}(t) + {\hat F}_-(t)%
\nonumber\\%
{\dot {\hat \sigma}}_{\mathrm{Z}}(t)&=&{\rm i} \frac{\Omega_R}{2}\,\left[{\hat \sigma}_+(t) - {\hat \sigma}_-(t)\right]
- \gamma \left[{\hat \sigma}_{\mathrm{Z}}(t) + \frac{1}{2} \right] + {\hat F}_\mathrm{Z}(t) \;.%
\nonumber\\%
\label{4.12}%
\end{eqnarray}
where $\Delta=\omega_c-\omega_0$, and $\gamma$ is the natural decay rate of the atomic excited state. The noise terms are given by
\begin{eqnarray}
{\hat F}_+(t)&=& + 2{\rm i} \frac{d_{0}}{\hbar}\,{\hat E}_{0z}^{(-)}(t)\,{\rm e}^{-{\rm i}\omega_c t}\,{\hat \sigma}_{\mathrm{Z}}(t)%
\nonumber\\%
{\hat F}_-(t)&=& - 2{\rm i} \frac{d_{0}}{\hbar}\,{\hat \sigma}_{\mathrm{Z}}(t)\,{\hat E}_{0z}^{(+)}(t)\,{\rm e}^{+{\rm i}\omega_c t}%
\nonumber\\%
{\hat F}_{\mathrm{Z}}(t)&=& + {\rm i}\frac{d_{0}}{\hbar}\left[\hat{\sigma}_+(t)\,{\hat E}_{0z}^{(+)} (t)\mathrm{e}^{+{\rm i}\omega_c t}  - \hat{E}_{0z}^{(-)}(t)\,\mathrm{e}^{-{\rm i} \omega t}\,\hat{\sigma}_-(t)\right]%
\nonumber\\%
\label{4.13}%
\end{eqnarray}
where the field operators $\hat{E}_{0z}^{(\pm)} (t)$ in (\ref{3.17}) are freely evolving Heisenberg images of the operators (\ref{a.7}). These equations are close-coupled but insufficient for a three-dimensional description. 

The weak excitation by the probe mode also concerns the satellite transitions $|a\rangle\to|x\rangle$ and $|a\rangle\to|y\rangle$, which evolve independently. They are similarly described and further we only follow the disturbed Heisenberg dynamics of creation/annihilation operators for $|a\rangle\to|x\rangle$ channel   
\begin{eqnarray}
\hat{\sigma}_{+}^{(x)}(t)&=&|x\rangle\langle a|(t)\,\mathrm{e}^{-i\omega_ct}%
\nonumber\\%
\hat{\sigma}_{-}^{(x)}(t)&=&|a\rangle\langle x|(t)\,\mathrm{e}^{+i\omega_ct}
\label{4.14}
\end{eqnarray}
The dynamics of $\hat{\sigma}_{+}^{(x)}(t)$ obey the following two coupled equations
\begin{eqnarray}
\lefteqn{\dot{\hat{\sigma}}_{+}^{(x)}(t)=\left(-{\rm i} \Delta - \frac{\gamma}{2}\right){\hat{\sigma}}_{+}^{(x)}(t) + \frac{\mathrm{i}}{2} \Omega_R\,|x\rangle\langle b|(t) +  {\hat F}_{+}^{(x)}(t)}%
\nonumber\\%
&&\frac{\partial}{\partial t}|x\rangle\langle b|(t)= - \gamma\,|x\rangle\langle b|(t) + \frac{\mathrm{i}}{2}\Omega_{R}\,{\hat{\sigma}}_{+}^{(x)}(t) + {\hat F}^{(xb)}(t)%
\nonumber\\%
\label{4.15}%
\end{eqnarray}
with the noise terms given by
\begin{eqnarray}
\hat{F}_{+}^{(x)}(t)&=&{\rm i} \frac{d_{0}}{\hbar}\,{\hat E}_{0z}^{(-)}(t)\,|x\rangle\langle b|(t)\,{\rm e}^{-{\rm i}\omega_c t}%
 \nonumber\\%
&&+ {\rm i} \frac{d_{0}}{\hbar}\,{\hat E}_{0x}^{(-)}(t)\,{\rm e}^{-{\rm i}\omega_c t}\,\big[|x\rangle\langle x|(t)-|a\rangle\langle a|(t)\big]%
\nonumber\\%
{\hat F}^{(xb)}(t)&=& + {\rm i}\frac{d_{0}}{\hbar}\left[\hat{\sigma}_{+}^{(x)}\!(t)\,{\hat E}_{0z}^{(+)}\!(t)\mathrm{e}^{+{\rm i}\omega_c t}\! -\!\hat{E}_{0x}^{(-)}\!(t)\,%
\mathrm{e}^{-{\rm i} \omega_c t}\,\hat{\sigma}_{-}\!(t)\right]%
\nonumber\\%
\label{4.16}
\end{eqnarray}
For the conjugated annihilation operator $\hat{\sigma}_{-}^{(x)}(t)$ the dynamics is driven by the equations Hermitian conjugated to (\ref{4.15}) and (\ref{4.16}).

The closed-coupled equations (\ref{4.12}) and (\ref{4.13}) are solved independently. But its solution further strongly affects the solution of equations (\ref{4.15}) and (\ref{4.16}) and is expressed in the dynamics of $\hat{\sigma}_{\pm}^{(x)}(t)$. That is a direct consequence of the fact that the system, being physically driven by Hamiltonian (\ref{3.6}), activates not only the occupied states, but the operator dynamics existing at the level of quantum fluctuations as well. Note that all these equations would simplify and become independent once the control field vanishes $\Omega_R\to 0$.

\subsubsection{The subject of calculation}

\noindent It is sufficient to evaluate only the positive frequency response of the susceptibility tensor (\ref{3.26}), which can be selected in the commutator expansions (\ref{4.3})-(\ref{4.5}). In the steady state regime for the dipole's response on a probe, polarized along principal axis $x$, such component is given by
\begin{equation}
\chi_{xx}^{(+)}(\tau)= \frac{i}{\hbar}n_0d_0^{2}\,\left\langle\left[\hat{\sigma}_{-}^{(x)}\!(t),\hat{\sigma}_{+}^{(x)}\!(t-\tau)\right]\right\rangle\,\mathrm{e}^{-i\omega_c\tau}%
\label{4.17}
\end{equation}
being independent on observation time $t$. Here, without loss of generality, we have assumed an infinite homogeneous medium with density $n_0$, and recall the comment after (\ref{3.18}). 

However the dipole's response on a probe, polarized along $z$-direction, is more tricky and consists of two contributions
\begin{equation}
\chi_{zz}^{(+)}(t,t')=\chi_{zz}^{(+-)}(t-t')+\chi_{zz}^{(++)}(t,t')\mathrm{e}^{+2i\mathbf{k}_c\cdot\mathbf{R}}%
\label{4.18}%
\end{equation}
where the first term expresses the dynamics of the atomic dipole driven by the positive frequency component of the probe mode. Similarly to (\ref{4.17}) it is given by
\begin{equation}
\chi_{zz}^{(+-)}(\tau)=\frac{i}{\hbar}n_0d_0^{2}\,\left\langle\left[\delta\hat{\sigma}_{-}^{\phantom{()}}\!(t),\delta\hat{\sigma}_{+}^{\phantom{()}}\!(t-\tau)\right]\right\rangle\,\mathrm{e}^{-i\omega_c\tau}%
\label{4.19}%
\end{equation}
where, for a sake of further derivation convenience, we have subtracted the mean expectation values of the operator functions and defined $\delta\hat{\sigma}_{\pm}\!(t)=\hat{\sigma}_{\pm}\!(t)-\bar{\sigma}_{\pm}\!(t)$.  The second term expresses the parametric response of the positive component driven by the negative frequency component of the probe
\begin{equation}
\left.\chi_{zz}^{(++)}\!(t,t')\right|_{t'=t-\tau}\!=\!%
\frac{i}{\hbar}n_0d_0^{2}\left\langle\left[\delta\hat{\sigma}_{-}^{\phantom{()}}\!(t),\delta\hat{\sigma}_{-}^{\phantom{()}}\!(t-\tau)\right]\right\rangle\mathrm{e}^{-i\omega_c(t+t')}%
\label{4.20}
\end{equation}
where the expectation value of the commutator is independent on the observation time $t$.

In the steady-state regime the susceptibility tensor is defined by the Fourier transformation of the above functions to the frequency representation. In the case considered, the spectral components are distributed in the vicinity of the reference frequencies $\omega\sim\omega_c,\omega_0$. 
Then we can parameterize the Fourier images by detuning $\Omega=\omega-\omega_c$ and apply it to the slow varying functions of $\tau$ contributing in (\ref{4.18})-(\ref{4.20}).

Omit the trivial multiplication factors as well as the exponents, structuring the spatial and temporal phase matching conditions, and define the following principal components of the susceptibility tensor:
\begin{equation}
\chi_{xx}^{(+)}(\Omega)\propto \int_0^{\infty}d\tau\, \mathrm{e}^{+i\Omega\tau}\,\left\langle\left[\hat{\sigma}_{-}^{(x)}\!(t),\hat{\sigma}_{+}^{(x)}\!(t-\tau)\right]\right\rangle%
\label{4.21}%
\end{equation}
and
\begin{equation}
\chi_{zz}^{(+-)}(\Omega)\propto \int_0^{\infty}d\tau\, \mathrm{e}^{+i\Omega\tau}\,\left\langle\left[\delta\hat{\sigma}_{-}\!(t),\delta\hat{\sigma}_{+}\!(t-\tau)\right]\right\rangle%
\label{4.22}%
\end{equation}
and
\begin{equation}
\chi_{zz}^{(++)}(\Omega)\propto \int_0^{\infty}d\tau\, \mathrm{e}^{-i\Omega\tau}\,\left\langle\left[\delta\hat{\sigma}_{-}\!(t),\delta\hat{\sigma}_{-}\!(t-\tau)\right]\right\rangle%
\label{4.23}%
\end{equation}
where we have paid attention that the retardation in (\ref{3.25}) dictates that $\tau>0$. 

Although the transformations (\ref{4.21})-(\ref{4.23}) look similar, there is an important difference in physics between them. The susceptibilities (\ref{4.21}) and (\ref{4.22}) are the elastic response of the probe mode $\omega=\omega_c+\Omega$ in the Rayleigh or stimulated quasi-Raman processes. In the latter case, the probe light is reconverted by stimulated scattering into the control mode and back to the probe. This type of coherent response is quite similar to those normally developing in an undisturbed dielectric medium. The critical difference is that in the nonlinear medium the probe experiences the quasi-energy structure, induced by the control field.  The susceptibility (\ref{4.23}) expresses a specific phase-sensitive nonlinear process, namely, the parametric conversion of the signal probe mode $\omega=\omega_{s}=\omega_c+\Omega$ to another idler probe mode $\omega_{i}=\omega_c-\Omega$. The spatial exponent in (\ref{4.18}) provides this process to expand primary under spatial phase matching conditions with $2\mathbf{k}_c\approx \mathbf{k}_s+\mathbf{k}_i$. That is fulfilled only approximately in an inhomogeneous sample.

We can explain our calculation algorithm of (\ref{4.21})-(\ref{4.23}) in an example of the $z$-polarized probe. Define the generalized Fourier transform for the operator fluctuations
\begin{eqnarray}
\delta\hat{\sigma}_{-}(\Omega)&=&\int_{-\frac{T}{2}}^{\frac{T}{2}}dt \; \mathrm{e}^{{\rm i}\Omega t} \delta\hat{\sigma}_{-}(t),%
\nonumber\\%
\delta\hat{\sigma}_{+}(\Omega)&=&\int_{-\frac{T}{2}}^{\frac{T}{2}}dt \; \mathrm{e}^{{\rm i}\Omega t} \delta\hat{\sigma}_{+}(t) = \delta\hat{\sigma}_{-}^\dagger(-\Omega),%
\nonumber\\%
\delta\hat{\sigma}_{\mathrm{Z}}(\Omega)&=&\int_{-\frac{T}{2}}^{\frac{T}{2}}dt \; \mathrm{e}^{{\rm i}\Omega t} \delta\hat{\sigma}_{\mathrm{Z}}(t),%
\label{4.24}%
\end{eqnarray}
which are parameterized by infinitely long $T\to +\infty$ and obey the periodic boundary conditions. This transforms equations (\ref{4.12}) to algebraic and analytically solvable equations. Then we make use of identities
\begin{eqnarray}
\lefteqn{\int_{-\infty}^{\infty} d\tau\,\mathrm{e}^{+{\rm i}\Omega \tau}\,\left\langle\left[\delta\hat{\sigma}_{-}(t),\delta\hat{\sigma}_{+}(t-\tau)\right] \right\rangle}
\nonumber\\%
&&=\lim_{T \to \infty} \frac{1}{T} \left\langle\left[ \delta\hat{\sigma}_{-}(\Omega),\delta\hat{\sigma}_{+}(-\Omega)\right] \right\rangle,
\label{4.25}%
\end{eqnarray}
and
\begin{eqnarray}
\lefteqn{\int_{-\infty}^{\infty} d\tau\,\mathrm{e}^{-{\rm i}\Omega \tau}\,\left\langle\left[\delta\hat{\sigma}_{-}(t),\delta\hat{\sigma}_{-}(t-\tau)\right] \right\rangle}
\nonumber\\%
&&=\lim_{T \to \infty} \frac{1}{T} \left\langle\left[ \delta\hat{\sigma}_{-}(-\Omega),\delta\hat{\sigma}_{-}(\Omega)\right] \right\rangle,
\label{4.26}%
\end{eqnarray}
That lets us construct the spectral expansion of the correlation functions (expectation values of the commutators in the frequency representation), formally containing both the retarded ($\tau>0$) and advanced ($\tau<0$) branches, and express them via correlation functions of the Langevin sources (\ref{4.13}), which we clarify below.

The retarded branch can be recovered by the following convolution transform of the Fourier images
\begin{equation}
\int_0^{\infty}d\tau\, \mathrm{e}^{\pm i\Omega\tau}\,f(\tau)=\mathrm{i}\int_{-\infty}^{+\infty}\frac{d\Omega'}{2\pi}\,\frac{F[f](\Omega')}{\pm\Omega-\Omega'+i0}%
\label{4.27}%
\end{equation}
where $F[f](\Omega')$ is the Fourier image of $f(\tau)$ in conventional form, i.e.
\begin{equation}
F[f](\Omega')=\int_{-\infty}^{+\infty}d\tau\,\mathrm{e}^{+{\rm i}\Omega'\tau}\,f(\tau)%
\label{4.28}%
\end{equation}
The integral in the right-hand side can be looped by a contour in any half-plane of the complex-valued  $\Omega'$ and expressed by sum of the residues of the integrand inside the contour.

\subsubsection{The correlation properties of the noise sources}

\noindent The described algorithm is crucially based on knowledge of the correlation properties of the noise terms. For equations (\ref{4.12}) the noise sources (\ref{4.13}) fulfill the following symmetry relations
\begin{equation}
\hat{F}_+^\dagger(t) = \hat{F}_-(t),\ \ \ \hat{F}_\mathrm{Z}^\dagger(t) = \hat{F}_\mathrm{Z}(t)%
\label{4.29}%
\end{equation}
and their correlations are relevantly approximated by delta-correlated Wiener-type random process with
\begin{equation}
\left\langle \hat{F}_q^\dagger(t)\hat{F}_{q^\prime}(t^\prime)\right\rangle = 2D_{qq^\prime}\,\delta(t - t^\prime)%
\label{4.30}
\end{equation}
where the diffusion coefficients can be structured in the following matrix
\begin{equation}
D_{qq^\prime} = \frac{\gamma}{2}\left(\begin{array}{ccc}
1 & 0 & \left\langle {\hat \sigma}_- (t) \right\rangle\\
0 & 0 & 0\\
\left\langle {\hat \sigma}_+ (t) \right\rangle & 0 & 1/2+\langle\hat{\sigma}_{\mathrm{Z}}(t)\rangle
\end{array}\right)%
\label{4.31}%
\end{equation}
with matrix raw/column ordered as $q,\,q^\prime = +,\,-,\, \mathrm{Z}$.  This matrix is assumed to be independent of time in the steady-state excitation regime. Then the mean values of $\langle\hat{\sigma}_{\pm}(t)\rangle$ and $\langle\hat{\sigma}_{Z}(t)\rangle$ are given by a stationary solution of the optical Bloch equations.

For equations (\ref{4.15}), (\ref{4.16}) and for their Hermitian counterparts we can order the correlation properties in form (\ref{4.30}) with the following matrix of diffusion coefficients
\begin{equation}
D_{qq^\prime} = \frac{\gamma}{2}\left(\begin{array}{cccc}
1 & 0 & \langle\hat{\sigma}_{-}(t)\rangle & 0 \\
0 & 0 & 0 & 0\\
\langle\hat{\sigma}_{+} (t)\rangle & 0 & 1/2+\langle\hat{\sigma}_{\mathrm{Z}}(t)\rangle & 0\\
0 & 0 & 0 & 0
\end{array}\right)%
\label{4.32}%
\end{equation}
where $q,\,q^\prime = {}_{+}^{(x)},\,{}_{-}^{(x)},\, {}^{(xb)},\,{}^{(bx)}$.

\vspace{\baselineskip}
\hrulefill
\vspace{\baselineskip}

\noindent By concluding this section, let us make one remark concerning the general description of the parametric process. Actually, the developed approach includes an optical nonlinearity incorporating all accessible orders of wave mixing. Since we have studied the nonlinear medium, originally driven by a single mode of signal light, the process generates only one phase-conjugated idler mode. If the signal light consisted of two modes, then the medium would respond in two idler modes as well, such that the four phase-conjugated modes together with the four photons of the control/pump field would be incorporated into an eight-wave mixing process. Furthermore, this option can be generalized up to any higher order of wave mixing.

\section{Results for a dilute system}\label{Section_V}

\noindent The solution of (\ref{4.12}) and (\ref{4.15}) is detailed in Appendix \ref{Appendix_B} and, in its general case, can be finalized only numerically. Here we present the principal components of the susceptibility tensor, calculated for different saturation parameters $s$, defined by (\ref{b.2}), and for $\Delta=0$ i.e. for the control field resonant to the reference atomic transition. The susceptibility components are scaled by a dimensionless factor expressed by the atomic density as
\begin{equation}
\frac{n_0\,d_0^{2}}{\hbar\gamma}=n_0\,\frac{3c^3}{4\omega_0^3} \equiv \frac{3}{4}n_0\,\lambdabar_0^3,
\label{5.1}%
\end{equation}
which for a dilute gas is a small quantity.  

\begin{figure}[pt]
\includegraphics[width=8.5cm]{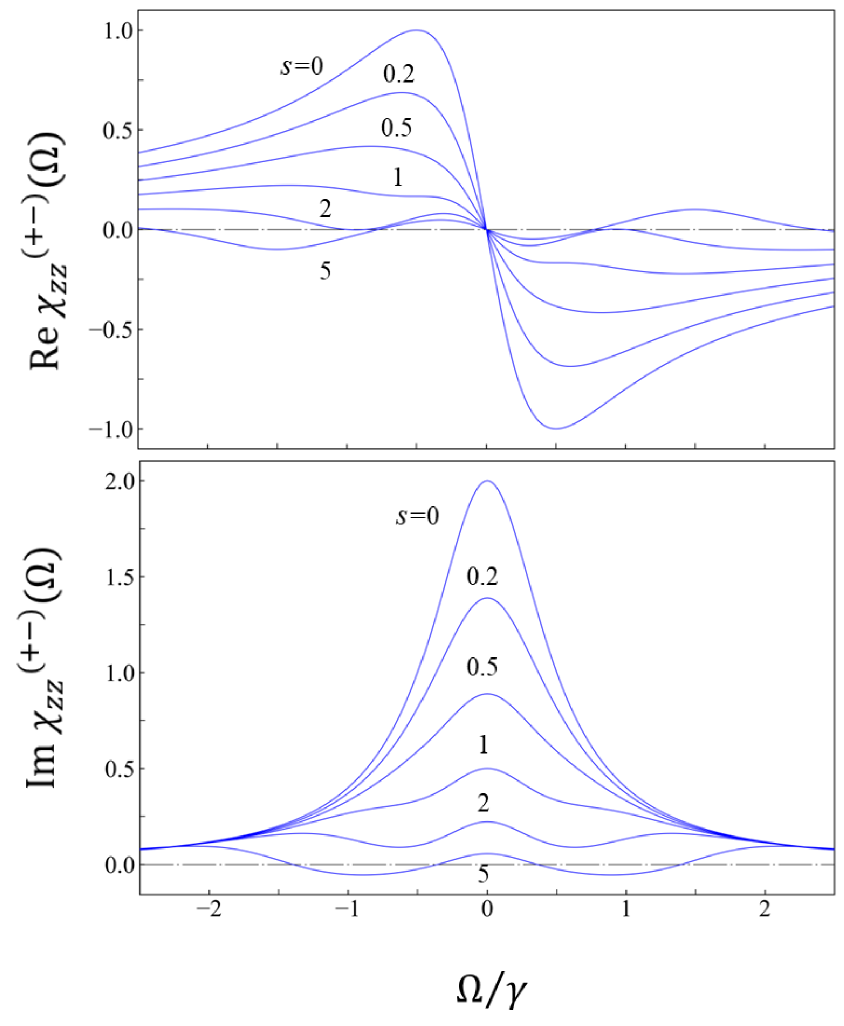}%
\caption{The Kerr-type nonlinear susceptibility for $z$-polarized probe and for the control field resonant to the atomic transition, and for different saturation parameters $s$.}
\label{fig3}%
\end{figure}

\begin{figure}[pt]
\includegraphics[width=8.5cm]{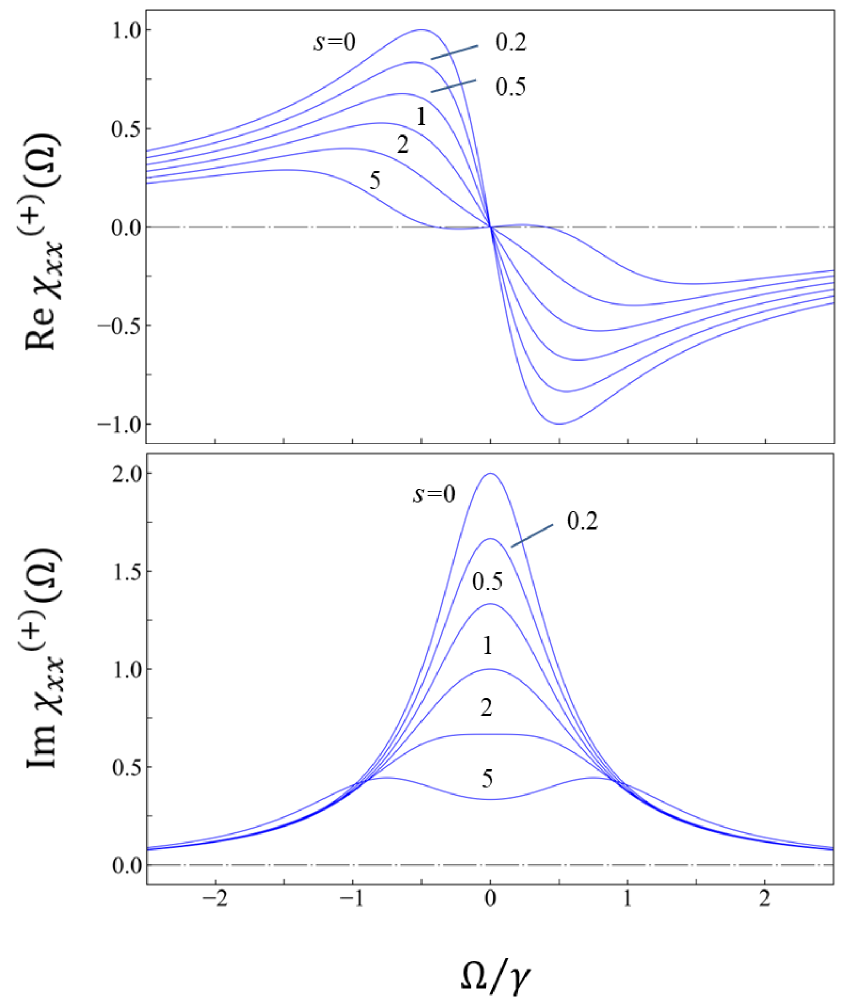}%
\caption{Same as in Fig.~\ref{fig3} but for the $x$-polarized probe.}
\label{fig4}%
\end{figure}

In Figs.~\ref{fig3} and \ref{fig4} we show the parts of the susceptibilily tensor responsible for the stimulated scattering and spontaneous losses of the probe modes, polarized respectively along the $z$ and $x$ directions. As pointed above, these processes are initiated by the Rayleigh and Raman-type scattering on the quasi-energy structure, distorted by the control field. Under vision of macroscopic Maxwell theory, such a medium response can be associated with the polarization current activated by the probe mode. In these processes, the positive frequency component of the polarization (transition) current is driven by the same positive frequency component of the probe. We shall refer to it as a general Kerr-type optical nonlinearity, where the conventional dielectric properties of the atomic medium are modified by the action of the strong control field. 

As follows from the dependencies of Fig.~\ref{fig3} the susceptibility of the $z$-polarized probe vanishes in the saturation limit when $s\gg 1$. However, the saturation regime is accompanied by a non-negligible manifestation of the negatively-valued imaginary part of the sample susceptibility leading to enhancement of the propagating light. That is a certain consequence for the probe light to be partly amplified by the spontaneous emission from the Mollow sidebands. Such an amplification mechanism can be utilized in an optically dense gas for preparation conditions of a random laser generation, as proposed earlier in \cite{Kaiser2009}.

The susceptibility component responding on the $x$-polarized probe does not vanish with enlarging $s$ but transforms to a doublet resonance structure, where the resonance maxima indicate the ac-Stark splitting of the ground state. As clarified in Appendix \ref{Appendix_B} in the saturation regime the atoms populate the ground state with one-half probability and experience the Autler-Townes splitting observable by the sample probe on the adjacent transitions in either $x$ or $y$ polarizations. A signature of the Autler-Townes resonance structure for $s\gg 1$ is foreseen from the dependencies in Fig.~\ref{fig4}.

\begin{figure}[t]
\includegraphics[width=8.5cm]{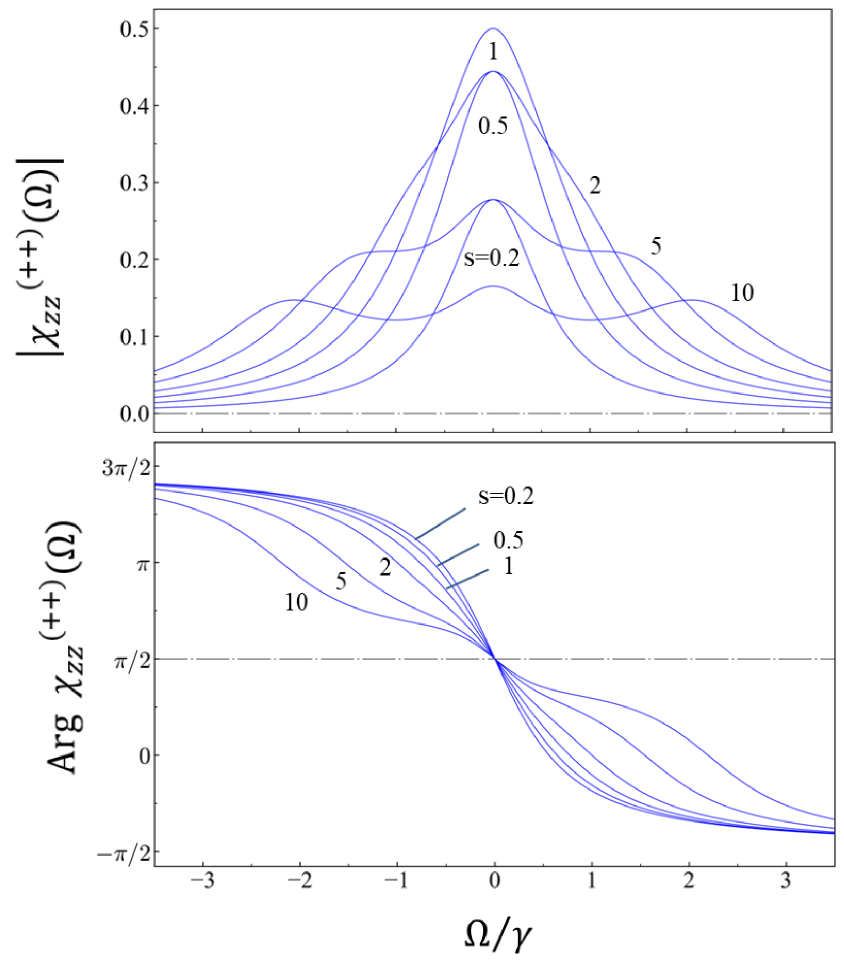}%
\caption{The parametric nonlinear susceptibility for the control field resonant to the atomic transition, and for different saturation parameters $s$. The upper and lower plots respectively show its absolute value and argument.}
\label{fig5}%
\end{figure}

The parametric coupling of the phase-conjugated modes, conventionally referred to as signal and idler, gives another type of nonlinearity. Let us focus on the optimal conditions for a maximally effective parametric amplification. The key feature of parametric nonlinearity is that the atomic medium is excluded from the energy interchange between the pump, signal, and idler modes. In other words, there are no energy losses in this interchange and the complexity of $\chi_{zz}^{(++)}(\Omega)$ indicates only the phase sensitivity of the parametric process. Unlike Figs.~\ref{fig3} and \ref{fig4}, instead of the real and imaginary parts, in Fig.~\ref{fig5} we show the absolute value and complex argument of the parametric nonlinear susceptibility responsible for the coupling of the phase-conjugated modes. This complex-valued quantity defines the coupling strength in an effective Hamiltonian governed by the joint dynamics of these modes under ideal conditions. 

The dependencies in the plots of Fig.~\ref{fig5} show that the efficiency of the parametric process cannot be infinitely enlarged if the pump becomes stronger. As follows from the calculated data, there is an optimal regime when the coupling strength is maximized near the intermediate saturation $s\sim 1$. For higher $s$ the spectrum becomes broader with the Mollow sidebands manifested. A local enhancement of the parametric coupling is observable near these sidebands.

\section{Interpolation to a high density}\label{Section_VI}

\noindent Let $\hat{T}^{(a)}$ be an arbitrary dyad-type operator associated with internal variables of an $a$-th atom of the atomic subsystem, introduced in Section \ref{Section_IV}. In general case its Heisenberg dynamics obeys the following equation
\begin{eqnarray}
\lefteqn{\dot{\hat{T}}^{(a)}(t)=\frac{i}{\hbar}\left[\hat{H}^{(a)\prime}_{\mathrm{Atom}}(t),\hat{T}^{(a)}(t)\right]}%
\nonumber\\%
&&\hspace{-0.5cm}-\frac{i}{\hbar}\left[\hat{d}_j^{(a)}(t),\hat{T}^{(a)}(t)\right]\,%
\frac{4\pi}{\cal V}\sum_{b\neq a}^{N}{}'\sum_{s}e\,_{sj}\,\left(\mathbf{e}_{s}\cdot\hat{\mathbf{d}}^{(b)}(t)\right)\,\mathrm{e}^{i\mathbf{k}_s\cdot(\mathbf{R}_a-\mathbf{R}_b)}%
\nonumber\\%
&&\hspace{-0.5cm}-\frac{i}{\hbar}\hat{E}_{\bot j}^{(-)}\!(\mathbf{R}_a,t)\left[\hat{d}_j^{(a)}\!(t),\hat{T}^{(a)}\!(t)\right]%
-\frac{i}{\hbar}\left[\hat{d}_j^{(a)}\!(t),\hat{T}^{(a)}\!(t)\right]\hat{E}_{\bot j}^{(+)}\!(\mathbf{R}_a,t)%
\nonumber\\%
\label{6.1}
\end{eqnarray}
where the structure of the atom-field interaction is clarified in Appendix \ref{Appendix_A}. The mode polarizations of the field vectors are defined in the Cartesian basis, and the default sum over repeated vector indices is assumed.

By substituting the displacement field (\ref{a.26}) into the interaction term, we have omitted the divergent contribution of the self-contact interaction. It is not voluntary, but, as shown in \cite{SKKH2009}, its action on the dipole dynamics is compensated by counteraction from the self-energy term, contributed to the atomic Hamiltonian; see (\ref{a.2})-(\ref{a.4}). So, the latter is also omitted in the atomic Hamiltonian of the particular $a$-th atom, which is pointed by a prime sign in (\ref{6.1}).

The field variables contribute to equations (\ref{6.1}) by the operators of transverse electric field expanded in its positive and negative frequency components
\begin{equation}
\hat{\mathbf{E}}_{\perp}(\mathbf{R},t)=\hat{\mathbf{E}}_{\perp}^{(+)}(\mathbf{R},t)+\hat{\mathbf{E}}_{\perp}^{(-)}(\mathbf{R},t)%
\label{6.2}
\end{equation}
taken at the point $\mathbf{R}=\mathbf{R}_a$ with running $a=1\ldots N$, where $N$ is the number of atoms. As commented in Appendix \ref{Appendix_A}, these operators do not commute with the atomic variables, and we have ordered them normally in equations (\ref{6.1}). There is no need in such an ordering for the dipole operators $\hat{\mathbf{d}}^{(b)}(t)$ since for $b\neq a$ and at coincident moments of time they always commute with $\hat{T}^{(a)}(t)$.

The operators (\ref{6.2}) fulfill the wave equation (\ref{2.6}), being a direct consequence of the Heisenberg-Maxwell equations (\ref{2.1}) or (\ref{2.4}). In (\ref{2.6}) the atomic subsystem affects the field dynamics via the collective polarization current, contributed to the right-hand side of (\ref{2.6}).  Thus, considered together, (\ref{6.2}) and (\ref{2.6}) give us a close-coupled system of the Heisenberg equations, describing the joint dynamics of the atoms and field in arbitrary external conditions and for any medium densities consistent with validity of the dipole long-wavelength approximation.

An issue naturally arises, could the system of equations (\ref{2.6}) and (\ref{6.1}) be statistically averaged and expressed by conventional form of the macroscopic Maxwell equations, written for the mean field and mean dipole polarization? If the atomic medium is dense and excited by a strong external coherent field, the answer is probably negative. That is because the Mollow nonlinear fluorescence would be emitted by the medium atoms in arbitrary directions and would interfere with an external probe mode and create large fluctuations of the local field amplitude. An attempt was made to describe the interference of such random waves, created by a coherent excitation in the saturation regime, specifically for a backward-scattering channel, in \cite{Delande2006a,Delande2006b}. The damaging role of fluctuations on coherent processes of light propagation through a disordered ultracold atomic gas was observed in experiments \cite{chaneliere2004,Balik2005,KSH2017}. 

In the considered case, if the coherent pump is as strong as it saturates the $|a\rangle\leftrightarrow|b\rangle$ optical transition, such field fluctuations can withdraw the option of four-wave mixing and parametric amplification. Nevertheless, the situation can be improved if the external coherent field is not extremely strong and induces relatively weak nonlinear disturbance to the system. Then, at zero approximation, equations (\ref{6.1}) and (\ref{2.6}) could be considered in linear regime and their correction by nonlinear dynamics could be taken into account as a perturbation to the light propagation process. In such an approximation the second term in (\ref{6.1}) reveals a local Lorentz-Lorenz correction to the electric field, added by proximal dipoles, and the linear permittivity of the medium can be recovered by construction of a self-consistent functional equation \cite{SKKH2009}. Its solution at the high density limit $n_0\lambdabar_0^3\gg 1$ shows a nontrivial spectral behavior in a vicinity of the atomic resonance. The permittivity can be negative so that the bulk atomic medium would have a forbidden zone preventing the penetration of a probe light inside the sample. 

Apparently, the parametric process can evolve in the spectral domain, where the medium possesses normal dielectric properties with a real-valued dielectric constant $\varepsilon\sim\varepsilon'>1$, having a vanishing imaginary part $\varepsilon''\to 0$. As shown in \cite{SKKH2009} in such a spectral domain, the radiation decay should be renormalized as $\gamma\to\gamma_{\varepsilon}=\sqrt{\epsilon}\gamma$. Any atomic dipole, existing as an exciton-type quasiparticle, would obey the dynamics similar to that described in Section \ref{Section_IV}, where the transverse electric field of the probe light, acting on the dipole, would be displaced by the Lorenz-Lorenz correction to the local field.

Thus we can see that the single atom response to a signal mode, constructed in Appendix \ref{Appendix_B}, is applicable for a dense atomic medium if (i) the control field is detuned to the domain, where the medium is transparent, (ii) the decay rate $\gamma$ is replaced by the renormalized decay constant $\gamma_{\varepsilon}$, (iii) and the solution, applied to a particular dipole, should be treated as a response on the local electric field. In such conditions, there is no physical resource for the parametric part of the susceptibility tensor to be significantly magnified by manipulating both the coherent pump and the sample density. In the considered model, the upper estimate for it cannot exceed the level of linear susceptibility of the sample in its transparency domain, evaluated in \cite{SKKH2009}.

\section{Conclusion}
\noindent In the paper, we have attempted to follow a rigorous microscopic description of the macroscopic phenomenon of optical nonlinearity. For this we have taken an example of a dense atomic ensemble consisting of neutral atoms interacting with an external electromagnetic field on a closed optical transition and framed our consideration by the long-wavelength dipole gauge. Although such a model deals with a rather simple matter object, it lets us enter to a more or less realistic physical visualization of a bulk medium, where the static interaction of proximal atomic dipoles interferes with their collective response on the transverse electric field, propagating through the medium and dressed by interaction with it. We were motivated to clarify whether there are any internal barriers in such an atomic medium that would prevent an unlimited magnification of optical nonlinear susceptibility.

In a low-density limit, for a dilute configuration, the system's response on a weak probe field is correctly described by the generalized Kubo formula, where the nonlinear action of the strong control field is incorporated to the system's steady-state Heisenberg-Langevin dynamics. Each atomic dipole transforms to a Mollow-type quasiparticle and a weak thermal disturbance converts the microscopic dynamics to relevancy of an average response of polarization current from mesoscopically scaled volumes of matter. 

The derived expression for the susceptibility tensor is naturally divided into a sum of the Kerr-type and parametric nonlinearities, the latter being highlighted by the phase-matching conditions. We have obtained that in the saturation regime, the coherent processes are suppressed and, as a consequence, the parametric response, induced by four-wave mixing of the pump (control) and signal (probe), has a maximum as function of the saturation parameter $s$ near the point $s\sim 1$. If the atomic medium were probed by two and more signal modes, then it would parametrically respond by the eight- and higher-order wave mixing processes, In any case, the parametric responses in any order, being coherent effects, would always be suppressed in the saturation regime for large $s\gg 1$.

It might be expected that the higher density of atoms would enhance the average mesoscopic polarization current and the parametric response in particular. However, the joint manifestation of the nonlinearity and density effects dramatically complicates the physical picture and makes the macroscopic description problematic. As pointed in Section \ref{Section_VI}, the randomly distributed Mollow nonlinear emission induces large fluctuations in the local electric field and breaks justification for a macroscopic description in terms of the mean electric and displacement fields. In this case, we have foreseen only the option to drive the dielectric medium by a relatively weak control field in those spectral domains, where the medium is transparent. In this case, the solution, originally obtained for a dilute configuration, could be renormalized and then would be applicable to a dense medium as well. 

Finally, we have obtained, as a key result, that the internal interactions and the excitation by a strong control field, taken as a joint effect, mainly modify the quasi-energy structure of the atomic medium and could not significantly magnify the coherent response of the polarization current on the driving fields. That can limit potential capabilities of quantum communication protocols, utilizing the entangled photon pairs, created by a parametric process, as a main resource of quantum correlations. As we think, these statements can be generalized and applicable for other bulk media, where interaction with the external field is provided by a polarization current localized in a mesoscopically scaled volume. In particular, referring to a security level of quantum communication protocols, the discussed physical bounds could limit potential interfering with an eavesdropper, in their exploiting the entangled photon pairs to attack of a quantum channel utilizing a two-photon or multi-photon component of the coherent field for quantum key distribution.

\section*{Acknowledgments} 
\noindent This work was supported by the Russian Science Foundation under Grant No. 23-72-10012, and by Rosatom in framework of the Roadmap for Quantum computing (Contracts No. 868-1.3-15/15-2021 and No. P2154). 
L.V.G. acknowledges support from the Foundation for the Advancement of Theoretical Physics and Mathematics “BASIS” under Grant No. 23-1-2-37-1. 

\subsection*{Conflict of interest}
\noindent The authors have no conflicts to disclose.

\appendix

\section{Microscopic description of the model}\label{Appendix_A}

\noindent For a dielectric medium, thinkable as a disordered atomic sample, the light-matter interaction is relevantly approached by a long-wavelength dipole-type interaction \cite{CohTann92}. Since the dipole gauge is only approximately valid, its practical use can lead to specific consequences, clarified by our derivation. That is mostly important for dense samples, where the atoms are separated by distances shorter than a scale of their radiation zones, purposing a correct description of near- and far-field interactions in the system dynamics.

The locality of the Maxwell theory lets us consider, as a basic process, a single-particle interaction with the field subsystem in one spatial point. Then we can straightforwardly generalize the results up to an arbitrary multiparticle system by superposing the partial contributions at the final derivation step. 

\subsection{Uncoupled Hamiltonian}

\noindent If the atoms are treated as neutral charge objects, having a small size, then by a long-wavelength cutoff of the field modes ($k\lesssim a_0^{-1}$, where $a_0$ is the Bohr radius) it makes possible to approximate interaction with the electromagnetic field by transformation from the Coulomb to the dipole gauge, see \cite{CohTann92}. That lets us formally follow the concept of a local point-like dipole interaction: (1) each atomic dipole is a compound object of vanishing size, (ii) the non-retarded scalar coupling of dipoles is eliminated by the gauge transformation so that each atom can interact only with the field and has retarded action on other atoms. 

Select an arbitrary atom of an ensemble, originate the frame with it, and follow its interaction with the field. The uncoupled Hamiltonian is divided in two contributions
\begin{equation}
\hat{H}_0=\hat{H}_{\mathrm{Atom}}+\hat{H}_{\mathrm{Field}}%
\label{a.1}
\end{equation}
The first term is Hamiltonian of the atom modified by dipole gauge as
\begin{equation}
\hat{H}_{\mathrm{Atom}}=\sum_{n}\hbar\omega_n|n\rangle\langle n|+\hat{H}_{\mathrm{self}}%
\label{a.2}%
\end{equation}
where $n$ is running over all the excited states and $\omega_n$ is a respective transition frequency counting from the ground state, and
\begin{eqnarray}
\hat{H}_{\mathrm{self}}&=&2\pi\int\frac{d^3 k}{(2\pi)^{3}}\left[\hat{\mathbf{d}}^2-(\mathbf{k}\cdot\hat{\mathbf{d}})\frac{(\mathbf{k}\cdot\hat{\mathbf{d}})}{k^2}\right]%
\nonumber\\%
&=& + \frac{4\pi}{3}\sum_{m}\,\hat{\mathbf{d}}^{2}\int\frac{d^3 k}{(2\pi)^{3}}= - \hat{\mathbf{d}}\cdot\hat{\mathbf{E}}_{\mathrm{dip}}(\mathbf{0})%
\nonumber\\%
\label{a.3}%
\end{eqnarray}
where $\hat{\mathbf{d}}$ is the operator of atomic dipole and
\begin{eqnarray}
\hat{\mathbf{E}}_{\mathrm{dip}}(\mathbf{R})&=&-4\pi\int\frac{d^3 k}{(2\pi)^{3}}\,\mathbf{k}\cdot\frac{{\mathbf{k}}\cdot\hat{\mathbf{d}}}{k^2}\;\mathrm{e}^{i\mathbf{k}\cdot\mathbf{R}}%
\nonumber\\
&=&-\nabla\,\hat{\phi}_{\mathrm{dip}}(\mathbf{R})%
\nonumber\\%
\hat{\phi}_{\mathrm{dip}}(\mathbf{R})&=&-4\pi\int\frac{d^3 k}{(2\pi)^{3}}\,\frac{i{\mathbf{k}}\cdot\hat{\mathbf{d}}}{k^2}\;\mathrm{e}^{i\mathbf{k}\cdot\mathbf{R}}=-\hat{\mathbf{d}}\cdot\nabla\frac{1}{R}%
\nonumber\\%
\label{a.4}%
\end{eqnarray}
expresses the static electric field of the dipole. The second term in (\ref{a.2}) introduces the dipole self-energy, infinitely divergent, and indicates inconsistency of the dipole gauge on a short spatial scale. Although this contribution seems nonphysical, it cannot be simply neglected.

The field contribution to the uncoupled Hamiltonian remains the same as in the original Coulomb gauge
\begin{equation}
\hat{H}=\sum_{s}\hbar\omega_s\,(a_s^{\dagger}a_s+1/2)%
\label{a.5}
\end{equation}
where sum over $s$ expands over all the field modes having frequencies $\omega_s$, where $a_s$ and $a_s^{\dagger}$ denote respectively the mode annihilation and creation operators  .

\subsection{Interaction}

\noindent The interaction terms formally reads
\begin{equation}
\hat{H}_{\mathrm{int}}= - \hat{\mathbf{d}}\cdot\hat{\mathbf{E}}(\mathbf{0})%
\label{a.6}
\end{equation}
To correctly interpret this term let us refer to the field transformation. There are the following three operators associated with the electric field. The operator
\begin{equation}
\hat{\mathbf{E}}(\mathbf{R})=\sum_s\left(\frac{2\pi\hbar\omega_s}{{\cal V}}\right)^{1/2}\left[i\mathbf{e}_s\,a_s\mathrm{e}^{+i\mathbf{k}\cdot\mathbf{R}}%
-i\mathbf{e}_s^\ast\,a_s^{\dagger}\mathrm{e}^{-i\mathbf{k}\cdot\mathbf{R}}\right]%
\label{a.7}%
\end{equation}
which, being original electric field operator in the Coulomb gauge, becomes the displacement field operator here. However, we leave its notation as $\hat{\mathbf{E}}$ in the case of a single dipole. 

The system is assumed to be loaded into a quantization box with volume ${\cal V}$ and mode specification $s=\mathbf{k},\sigma$, where $\sigma=1,2$ for two orthogonal polarizations $\mathbf{e}_s\equiv\mathbf{e}_\sigma(\mathbf{k})$. The operator of displacement field can be expanded in sum of the transverse electric field and transverse component of the atomic dipole operators as follows
\begin{eqnarray}
\hat{\mathbf{E}}(\mathbf{R})&=&\hat{\mathbf{E}}_{\perp}(\mathbf{R})+\frac{4\pi}{\cal V}\sum_{s}\mathbf{e}_s\cdot(\mathbf{e}_s\cdot \hat{\mathbf{d}})\,\mathrm{e}^{i\mathbf{k}\cdot\mathbf{R}}%
\nonumber\\%
\hat{\mathbf{E}}(\mathbf{R})&\equiv&\hat{\mathbf{E}}_{\perp}(\mathbf{R})+ 4\pi\,\hat{\mathbf{d}}_{\perp}(\mathbf{R})%
\label{a.8}%
\end{eqnarray}
where for a sake of simplicity we have implied the real basis vectors: $\mathbf{e}_s=\mathbf{e}_s^{\ast}$. The complete electric field is given by the operator
\begin{eqnarray}
\hat{\mathbf{E}}_{\mathrm{tot}}(\mathbf{R})&=&\hat{\mathbf{E}}(\mathbf{R})-4\pi\,\hat{\mathbf{d}}\,\delta(\mathbf{R})%
\nonumber\\%
&=&\hat{\mathbf{E}}_{\perp}(\mathbf{R})+\hat{\mathbf{E}}_{\mathrm{dip}}(\mathbf{R})%
\label{a.9}%
\end{eqnarray}
with the second term in the second line defined by (\ref{a.4}). This clarifies the substance of dipole approximation where the longitudinal field is only approximately approached by the dipoles's static field.

The second term in (\ref{a.8}) creates the self-contact interaction. Naively, we arrive at the contradictory situation. The substitution of expansion (\ref{a.8}) into (\ref{a.6}) would add an extra divergent term, having the same physical nature as (\ref{a.3}). However, we prevent substituting the terms, selected by (\ref{a.8}), into the system Hamiltonian separately. The dipole's canonic operators $\hat{\mathbf{r}}$ and $-i\hbar\nabla$ commute with the complete displacement field operator (\ref{a.7}) as well as with its frequency components, but they do not commute with its partial contributions selected by (\ref{a.8}).

In fact, as explained in \cite{SKKH2009}, the joint action of two divergent contributions -- dipolar self-energy (\ref{a.3}) and the contact interaction coming from (\ref{a.8}) -- compensate each other and do not affect the dipole's dynamics. Eventually for a many-particle system we have to follow the derivation algorithm of that paper, which is intrinsically based on the conventional arguments of the macroscopic Maxwell approach. Expansion (\ref{a.8}) introduces a physically clear dynamical picture. The proximal dipoles (separated by distances within their radiation zones) are indistinguishable and coupled into a mesoscopic cluster of local polarization current, and they respond on the external driving field cooperatively. In such conditions the transverse electric field, being thermally averaged, will propagate along the sample as an electromagnetic wave mediated by the macroscopic Maxwell equations with retardation.

\subsection{Derivation of the Maxwell-Heisenberg equations}

\noindent Let us construct the Heisenberg equations for the field subsystem. That part of the system Hamiltonian, which overlaps the field variables with atomic subsystem, in the dipole gauge is given by
\begin{equation}
\hat{H}=\sum_{s}\hbar\omega_s\,(a_s^{\dagger}a_s+1/2) - \hat{\mathbf{d}}\cdot\hat{\mathbf{E}}(\mathbf{0})+...%
\label{a.10}
\end{equation}
The Heisenberg equation for the time dependent operator of the displacement field reads
\begin{equation}
\dot{\hat{\mathbf{E}}}(\mathbf{R},t)=\frac{i}{\hbar}\left[\hat{H}(t),\,\hat{\mathbf{E}}(\mathbf{R},t)\right]%
\label{a.11}%
\end{equation}
where in the right hand side we have extended the time independent Schr\"odinger commutator of the operators (\ref{a.10}) and (\ref{a.7}) to its time-dependent Heisenberg form. Throughout this section, while evaluating any of the commutators, we will follow the Schr\"odinger representation and add the Heisenberg time dependence to the final results.

The commutator with the free Hamiltonian leads
\begin{eqnarray}
\lefteqn{\left[\sum_{s}\hbar\omega_s\,(a_s^{\dagger}a_s+1/2),\,\hat{\mathbf{E}}(\mathbf{R})\right]}
\nonumber\\%
&&=\sum_s\left(\frac{2\pi\hbar\omega_s}{{\cal V}}\right)^{1/2}\left[-i\hbar\omega_s\mathbf{e}_s\,a_s\mathrm{e}^{+i\mathbf{k}\cdot\mathbf{R}}%
-i\hbar\omega_s\mathbf{e}_s^\ast\,a_s^{\dagger}\mathrm{e}^{-i\mathbf{k}\cdot\mathbf{R}}\right]%
\nonumber\\%
\label{a.12}%
\end{eqnarray}
where the right-hand side can be expressed via magnetic field. Indeed, we have
\begin{eqnarray}
\lefteqn{\hat{\mathbf{A}}(\mathbf{R})=\sum_s\left(\frac{2\pi\hbar c^2}{{\omega_s\cal V}}\right)^{1/2}\left[\mathbf{e}_s\,a_s\mathrm{e}^{+i\mathbf{k}\cdot\mathbf{R}}%
+\mathbf{e}_s^\ast\,a_s^{\dagger}\mathrm{e}^{-i\mathbf{k}\cdot\mathbf{R}}\right]}%
\nonumber\\%
&&\hat{\mathbf{B}}(\mathbf{R})=\mathrm{rot}\,\hat{\mathbf{A}}(\mathbf{R})%
\nonumber\\%
&&\sum_s\left(\frac{2\pi\hbar c^2}{{\omega_s\cal V}}\right)^{1/2}\left[i[\mathbf{k}\times\mathbf{e}_s]\,a_s\mathrm{e}^{+i\mathbf{k}\cdot\mathbf{R}}%
-i[\mathbf{k}\times\mathbf{e}_s^\ast]\,a_s^{\dagger}\mathrm{e}^{-i\mathbf{k}\cdot\mathbf{R}}\right]%
\nonumber\\%
\label{a.13}%
\end{eqnarray}
and (\ref{a.12}) can be rewritten as
\begin{equation}
\left[\sum_{s}\hbar\omega_s\,(a_s^{\dagger}a_s+1/2),\,\hat{\mathbf{E}}(\mathbf{R})\right]=-i\hbar\,c\,\mathrm{rot}\,\hat{\mathbf{B}}(\mathbf{R})%
\label{a.14}
\end{equation}
where we have made use of the identity $[\mathbf{k}\times[\mathbf{k}\times\mathbf{e}_s]]=-\mathbf{k}^2\mathbf{e}_s=-\omega_s^2/c^2\,\mathbf{e}_s$.

The commutator with the second term of (\ref{a.10}) is proportional to
\begin{eqnarray}
\lefteqn{\left[\hat{E}_{\mu}(\mathbf{0}),\,\hat{E}_{\nu}(\mathbf{R})\right]=\sum_s\left(\frac{2\pi\hbar\omega_s}{{\cal V}}\right)}%
\nonumber\\%
&&\times\left[+i(-i)e_{s\mu}e_{s\nu}^\ast\,[a_s,a_s^{\dagger}]\mathrm{e}^{-i\mathbf{k}\cdot\mathbf{R}}%
-i(+i)e_{s\mu}^\ast e_{s\nu}\,[a_s^\dagger,a_s]\mathrm{e}^{+i\mathbf{k}\cdot\mathbf{R}}\right]%
\nonumber\\%
\nonumber\\%
&&=\sum_s\left(\frac{2\pi\hbar\omega_s}{{\cal V}}\right)\left[e_{s\mu}e_{s\nu}^\ast\,\mathrm{e}^{-i\mathbf{k}\cdot\mathbf{R}}%
-e_{s\mu}^\ast e_{s\nu}\,\mathrm{e}^{+i\mathbf{k}\cdot\mathbf{R}}\right]%
\label{a.15}%
\end{eqnarray}
For a sake of the derivation simplicity we can imply the real basis vectors: $\mathbf{e}_s=\mathbf{e}_s^{\ast}$. Then, after changing $\mathbf{k}\to-\mathbf{k}$ in the second term, in the square brackets we arrive at
\begin{eqnarray}
\left[\hat{E}_{\mu}(\mathbf{0}),\,\hat{E}_{\nu}(\mathbf{R})\right]&=&0%
\nonumber\\%
\left[-\,\hat{\mathbf{d}}\cdot\hat{\mathbf{E}}(\mathbf{0}),\,\hat{\mathbf{E}}(\mathbf{R})\right]&=&0%
\label{a.16}%
\end{eqnarray}
i.e. the operator of displacement field commutes with the interaction part of the system Hamiltonian.

The Heisenberg equation for the time dependent operator of the magnetic field reads
\begin{equation}
\dot{\hat{\mathbf{B}}}(\mathbf{R},t)=\frac{i}{\hbar}\left[\hat{H}(t),\,\hat{\mathbf{B}}(\mathbf{R},t)\right]%
\label{a.17}%
\end{equation}
Then by disclosing the right hand side for the operators in the Schr\"odinger representation we subsequently obtain
\begin{eqnarray}
\lefteqn{\left[\sum_{s}\hbar\omega_s\,(a_s^{\dagger}a_s+1/2),\,\hat{\mathbf{B}}(\mathbf{R})\right]}
\nonumber\\%
&=&\sum_s\left(\frac{2\pi\hbar\omega_s}{{\cal V}}\right)^{1/2}\left[-i\hbar\omega_s[\mathbf{k}\times\mathbf{e}_s]\,a_s\mathrm{e}^{+i\mathbf{k}\cdot\mathbf{R}}\right.%
\nonumber\\%
&&\phantom{=\sum_s}\left.-i\hbar\omega_s[\mathbf{k}\times\mathbf{e}_s^\ast]\,a_s^{\dagger}\mathrm{e}^{-i\mathbf{k}\cdot\mathbf{R}}\right]%
\nonumber\\%
&=&i\hbar\,c\,\mathrm{rot}\,{\hat{\mathbf{E}}(\mathbf{R})}
\label{a.18}%
\end{eqnarray}
and for the interaction part, similarly to (\ref{a.15}), we have to evaluate the commutator
\begin{eqnarray}
\lefteqn{\left[\hat{E}_{\mu}(\mathbf{0}),\,\hat{B}_{\nu}(\mathbf{R})\right]=\sum_s\left(\frac{2\pi\hbar\,c}{{\cal V}}\right)}%
\nonumber\\%
&&\times\left[+i(-i)e_{s\mu}[\mathbf{k}\times\mathbf{e}_s^\ast]_\nu\,[a_s,a_s^{\dagger}]\mathrm{e}^{-i\mathbf{k}\cdot\mathbf{R}}\right.%
\nonumber\\%
\nonumber\\%
&&\left.-i(+i)e_{s\mu}^\ast [\mathbf{k}\times\mathbf{e}_s]_{\nu}\,[a_s^\dagger,a_s]\mathrm{e}^{+i\mathbf{k}\cdot\mathbf{R}}\right]%
\nonumber\\%
\label{a.19}%
\end{eqnarray}
We further obtain
\begin{eqnarray}
\lefteqn{\left[-\,\hat{\mathbf{d}}\cdot\hat{\mathbf{E}}(\mathbf{0}),\,\hat{\mathbf{B}}(\mathbf{R})\right]= -\sum_s\left(\frac{2\pi\hbar\,c}{{\cal V}}\right)}%
\nonumber\\%
&&\times\left[(\hat{\mathbf{d}}\cdot\mathbf{e}_{s})[\mathbf{k}\times\mathbf{e}_s^\ast]\,\mathrm{e}^{-i\mathbf{k}\cdot\mathbf{R}}%
-(\hat{\mathbf{d}}\cdot\mathbf{e}_{s}^\ast) [\mathbf{k}\times\mathbf{e}_s]\,\mathrm{e}^{+i\mathbf{k}\cdot\mathbf{R}}\right]%
\nonumber\\%
\nonumber\\%
&&=-\,i\hbar\,c\,\mathrm{rot}\sum_s\frac{2\pi}{{\cal V}}%
\left[(\hat{\mathbf{d}}\cdot\mathbf{e}_{s})\,\mathbf{e}_s^\ast\,\mathrm{e}^{-i\mathbf{k}\cdot\mathbf{R}}%
+(\hat{\mathbf{d}}\cdot\mathbf{e}_{s}^\ast)\,\mathbf{e}_s\,\mathrm{e}^{+i\mathbf{k}\cdot\mathbf{R}}\right]%
\nonumber\\%
&&=-\,i\hbar\,4\pi\,c\,\mathrm{rot}\int\frac{d^3k}{(2\pi)^3}\left[\hat{\mathbf{d}}-\mathbf{k}\cdot\frac{{\mathbf{k}}\cdot\hat{\mathbf{d}}}{k^2}\right]\,\mathrm{e}^{+i\mathbf{k}\cdot\mathbf{R}}%
\nonumber\\%
\nonumber\\%
&&=-\,i\hbar\,c\,\mathrm{rot}\,\left[4\pi\,\hat{\mathbf{d}}_{\perp}(\mathbf{R})\right]=-\,i\hbar\,c\,\mathrm{rot}\,\left[4\pi\,\hat{\mathbf{d}}\,\delta(\mathbf{R})\right]%
\nonumber\\%
\label{a.20}%
\end{eqnarray}
where we have applied some obvious transformation rules.

Eventually (\ref{a.14}), (\ref{a.16}), (\ref{a.18}), (\ref{a.20}) lead us to the Maxwell equations in the following form
\begin{eqnarray}
\mathrm{rot}\,\hat{\mathbf{B}}(\mathbf{R},t)&=&\frac{1}{c}\,\dot{\hat{\mathbf{E}}}(\mathbf{R},t)%
\nonumber\\%
\mathrm{rot}\,\left[\hat{\mathbf{E}}(\mathbf{R},t)-4\pi\,\hat{\mathbf{d}}_{\perp}(\mathbf{R},t)\right]&=&-\frac{1}{c}\,\dot{\hat{\mathbf{B}}}(\mathbf{R},t)%
\label{a.21}
\end{eqnarray}
or in the equivalent form
\begin{eqnarray}
\mathrm{rot}\,\hat{\mathbf{B}}(\mathbf{R},t)&=&\frac{1}{c}\,\dot{\hat{\mathbf{E}}}_{\perp}(\mathbf{R},t)+\frac{4\pi}{c}\,\dot{\hat{\mathbf{d}}}_{\perp}(\mathbf{R},t)%
\nonumber\\%
\mathrm{rot}\,\hat{\mathbf{E}}_{\perp}(\mathbf{R},t)&=&-\frac{1}{c}\,\dot{\hat{\mathbf{B}}}(\mathbf{R},t)%
\label{a.22}%
\end{eqnarray}
and we can construct the closed equation for the transverse electric field
\begin{equation}
\triangle \hat{\mathbf{E}}_{\perp}(\mathbf{R},t)-\frac{1}{c^2}\ddot{\hat{\mathbf{E}}}_{\perp}(\mathbf{R},t)=\frac{4\pi}{c^2}\,\ddot{\hat{\mathbf{d}}}_{\perp}(\mathbf{R},t)%
\label{a.23}%
\end{equation}
which we have to consider as coupled with the complementary Heisenberg equations written for all the atomic dipoles.

Due to locality the Maxwell equations (\ref{a.22}) can be straightforwardly generalized for a multi-particle system, by constructing the sum over the dipoles, contributing to the right-hand side in other spatial points. Furthermore the longitudinal field can be added in the second line of (\ref{a.22}). Thus in the general case we arrive at
\begin{eqnarray}
\mathrm{rot}\,\hat{\mathbf{B}}(\mathbf{R},t)&=&\frac{1}{c}\,\dot{\hat{\mathbf{E}}}_{\perp}(\mathbf{R},t)+\frac{4\pi}{c}\,\sum_{a=1}^{N}\dot{\hat{\mathbf{d}}}_{\perp}^{(a)}(\mathbf{R},t)\equiv\dot{\hat{\mathbf{D}}}(\mathbf{R},t)%
\nonumber\\%
\mathrm{rot}\,\hat{\mathbf{E}}_{\mathrm{tot}}(\mathbf{R},t)&=&-\frac{1}{c}\,\dot{\hat{\mathbf{B}}}(\mathbf{R},t)%
\label{a.24}%
\end{eqnarray}
where we have generalized (\ref{a.8}) and (\ref{a.9}) for the multi-particle system, and defined in the Schr\"{o}dinger picture
\begin{equation}
\hat{\mathbf{d}}_{\perp}^{(a)}(\mathbf{R})=\frac{1}{\cal V}\sum_{s}\mathbf{e}_s\cdot(\mathbf{e}_s\cdot \hat{\mathbf{d}}^{(a)})\,\mathrm{e}^{i\mathbf{k}\cdot(\mathbf{R}-\mathbf{R}_a)}%
\label{a.25}
\end{equation}
where $\mathbf{R}_a$ is the spatial location of the $a$-th dipole, and the fields operators are given by
\begin{eqnarray}
\hat{\mathbf{D}}(\mathbf{R})&=&\hat{\mathbf{E}}_{\perp}(\mathbf{R})+ 4\pi\,\sum_{a=1}^{N}\hat{\mathbf{d}}_{\perp}^{(a)}(\mathbf{R})%
\nonumber\\%
\hat{\mathbf{E}}_{\mathrm{tot}}(\mathbf{R})&=&\hat{\mathbf{E}}_{\perp}(\mathbf{R})+\sum_{a=1}^{N}\hat{\mathbf{E}}_{\mathrm{dip}}^{(a)}(\mathbf{R})%
\label{a.26}%
\end{eqnarray}
with the longitudinal field component constructed by sum of partial contributions from each dipole, see (\ref{2.4})
\begin{eqnarray}
\hat{\mathbf{E}}_{\mathrm{dip}}^{(a)}(\mathbf{R})&=&-4\pi\int\frac{d^3 k}{(2\pi)^{3}}\,\mathbf{k}\cdot\frac{{\mathbf{k}}\cdot\hat{\mathbf{d}}^{(a)}}{k^2}\;\mathrm{e}^{i\mathbf{k}\cdot(\mathbf{R}-\mathbf{R}_a)}%
\nonumber\\
&=&-\nabla\,\hat{\phi}_{\mathrm{dip}}^{(a)}(\mathbf{R})%
\nonumber\\%
\hat{\phi}_{\mathrm{dip}}^{(a)}(\mathbf{R})&=&-4\pi\int\frac{d^3 k}{(2\pi)^{3}}\,\frac{i{\mathbf{k}}\cdot\hat{\mathbf{d}}^{(a)}}{k^2}\;\mathrm{e}^{i\mathbf{k}\cdot(\mathbf{R}-\mathbf{R}_a)}%
\nonumber\\%
&=&-\hat{\mathbf{d}}^{(a)}\cdot\nabla\frac{1}{|\mathbf{R}-\mathbf{R}_a|}%
\label{a.27}%
\end{eqnarray}
Equations (\ref{a.24})-(\ref{a.27}) reproduce the microscopic background of the macroscopically structured Maxwell-Heisenberg equations used in the main text.

\section{Evaluation of the susceptibility tensor}\label{Appendix_B}

\noindent In steady state regime the solution of (\ref{4.12}) for mean and time independent values of atomic operators (\ref{4.11}) is given by
\begin{eqnarray}
\langle\sigma_{-}\rangle&\equiv&\bar{\sigma}_{-}=-\frac{\Delta-i\displaystyle{\frac{\gamma}{2}}}{\Omega_R}\frac{s}{s+1}%
\nonumber\\%
\langle\sigma_{+}\rangle&\equiv&\bar{\sigma}_{+}=-\frac{\Delta+i\displaystyle{\frac{\gamma}{2}}}{\Omega_R}\frac{s}{s+1}%
\nonumber\\%
\langle\sigma_{Z}\rangle&\equiv&\bar{\sigma}_{Z}=-\frac{1}{2}\,\frac{1}{s+1}%
\label{b.1}
\end{eqnarray}
where
\begin{equation}
s=\frac{\Omega_R^2}{2}\frac{1}{\Delta^2+\displaystyle{\frac{\gamma^2}{4}}}%
\label{b.2}
\end{equation}
is the so named saturation parameter varied between $s\ll 1$ (weak nonlinearity) to $s\gg 1$ (saturation limit). These basic relations, being, in fact, the steady state solution of optical Bloch equations, are further used in parameterization of the Heisenberg-Langevin stochastic dynamics via correlation properties of the noise sources (\ref{4.31}) and (\ref{4.32}).

\subsection{Steady-state stochastic dynamics of (\ref{4.12}), (\ref{4.15})}

\noindent With Fourier transform (\ref{4.24}) (together with similar transformation for other Heisnberg components of the atomic operators as well as of the noise sources) applied to (\ref{4.12}) and (\ref{4.15}) we arrive at the following sets of algebraic equations for the Fourier images
\begin{eqnarray}
\left(\Delta-\Omega - i\frac{\gamma}{2}\right)\delta{\hat{\sigma}}_{+}(\Omega)-\Omega_R\,\delta\hat{\sigma}_{\mathrm{Z}}(\Omega) &=& -i\,\hat{F}_{+}(\Omega)%
\nonumber\\%
\left(\Delta+\Omega + i\frac{\gamma}{2}\right)\delta{\hat{\sigma}}_{-}(\Omega)-\Omega_R\,\delta\hat{\sigma}_{\mathrm{Z}}(\Omega) &=& +i\,\hat{F}_{-}(\Omega)%
\nonumber\\%
\left(\Omega+i\gamma\right){\hat{\sigma}}_{\mathrm{Z}}(\Omega)+\frac{\Omega_R}{2}\,\left[\delta\hat{\sigma}_{+}(\Omega) - \delta\hat{\sigma}_{-}(\Omega)\right]&=&i\,\hat{F}_\mathrm{Z}(\Omega)%
\nonumber\\%
\label{b.3}%
\end{eqnarray}
and
\begin{eqnarray}
\lefteqn{\left(\Delta+\Omega-i\frac{\gamma}{2}\right)\hat{\sigma}_{+}^{(x)}(-\Omega) - \frac{1}{2} \Omega_R\,|x\rangle\langle b|(-\Omega) = -i  {\hat F}_{+}^{(x)}(-\Omega)}%
\nonumber\\%
&&(\Omega-i\gamma)|x\rangle\langle b|(-\Omega)-\frac{1}{2}\Omega_{R}\,{\hat{\sigma}}_{+}^{(x)}(-\Omega) = -i {\hat F}^{(xb)}(-\Omega)%
\nonumber\\%
\label{b.4}%
\end{eqnarray}
where we have made used of the periodic boundary conditions for integration limits in (\ref{4.24}) in the regime of stationary fluctuations.

These equations can be resolved straightforwardly, and the expectation values of the commutators of atomic variables can be expressed by the commutators of noise fluctuations.  The latter satisfy (4.29) - (4.32) and, because of their delta-type temporal correlations, have a flat spectral density, parameterized by the components of the density matrix (\ref{b.1}). We omit here the quite cumbersome derivation sequence and present only the final results. 

Finally, with recovering the dimension factors in basic definitions (\ref{4.21})-(\ref{4.23}), the principal components of the susceptibility tensor are given by the following integral expansions 
\begin{equation}
\chi_{zz}^{(+-)}(\Omega)=-\frac{1}{\hbar}n_0d_0^2\int\limits_{-\infty}^{+\infty}\!\frac{d\Omega'}{2\pi}\,\frac{1}{\Omega-\Omega'+i0}\,\frac{N^{(+-)}(\Omega')}{D^{(+-)}(\Omega')},%
\label{b.5}%
\end{equation}
and
\begin{equation}
\chi_{zz}^{(++)}(\Omega)=-\frac{1}{\hbar}n_0d_0^2\int\limits_{-\infty}^{+\infty}\!\frac{d\Omega'}{2\pi}\,\frac{1}{-\Omega-\Omega'+i0}\,\frac{N^{(++)}(\Omega')}{D^{(++)}(\Omega')}%
\label{b.6}%
\end{equation}
and
\begin{equation}
\chi_{xx}^{(+)}(\Omega)=\chi_{yy}^{(+)}(\Omega)=-\frac{1}{\hbar}n_0d_0^2\int\limits_{-\infty}^{+\infty}\!\frac{d\Omega'}{2\pi}\,\frac{1}{\Omega-\Omega'+i0}\,\frac{N_{\perp}(\Omega')}{D_{\perp}(\Omega')},%
\label{b.7}%
\end{equation}
where the $z$-component is split in two parts: the elastic Kerr-type coherent response $\chi_{zz}^{(+-)}(\Omega)$ and parametric part $\chi_{zz}^{(++)}(\Omega)$ responsible for creation of a phase conjugated idler mode.

The denominators under integrands in (\ref{b.5}) and (\ref{b.6}) are given by
\begin{equation}
D^{(+-)}(\Omega)\equiv\left[\left(\Delta+\Omega\right)^2+\displaystyle{\frac{\gamma^2}{4}}\right]\,|M(\Omega;\gamma,\Delta,\Omega_R)|^2%
\label{b.8}
\end{equation}
and 
\begin{equation}
D^{(++)}(\Omega)\equiv\left(\Delta+\Omega+i\displaystyle{\frac{\gamma}{2}}\right)\left(\Delta-\Omega+i\displaystyle{\frac{\gamma}{2}}\right)\,|M(\Omega;\gamma,\Delta,\Omega_R)|^2%
\label{b.9}
\end{equation}
where 
\begin{widetext}
\begin{equation}
M(\Omega;\gamma,\Delta,\Omega_R)=i\displaystyle{\frac{\gamma}{2}}\left[\Delta^2-\left(\Omega + i\displaystyle{\frac{\gamma}{2}}\right)^2\right]%
+(\Omega+i\displaystyle{\frac{\gamma}{2}})\left[\Delta^2-\left(\Omega + i\displaystyle{\frac{\gamma}{2}}\right)^2+\Omega_R^2\right]
\label{b.10}
\end{equation}
is a cubic polynomial of $\Omega$ that defines the locations of the Mollow-triplet resonances.

The numerators under integrands in (\ref{b.5}) and (\ref{b.6}) are given by
\begin{eqnarray}
N^{(+-)}(\Omega)&=&\gamma\,\left|M(\Omega;\gamma,\Delta,\Omega_R)\right|^2-\gamma\,\Omega_R^4\,\Delta\,\Omega%
+\frac{\gamma}{2}\,\Omega_R^2\left(\Delta-\Omega-i\displaystyle{\frac{\gamma}{2}}\right)M^{\ast}(\Omega;\gamma,\Delta,\Omega_R)%
+\frac{\gamma}{2}\,\Omega_R^2\left(\Delta-\Omega + i\displaystyle{\frac{\gamma}{2}}\right)M(\Omega;\gamma,\Delta,\Omega_R)%
\nonumber\\%
\nonumber\\%
&&+\gamma\bar{\sigma}_{-}\cdot\Omega_R\left(\Delta+\Omega-i\displaystyle{\frac{\gamma}{2}}\right)%
\left[\left(\Delta-\Omega+i\displaystyle{\frac{\gamma}{2}}\right)M(\Omega;\gamma,\Delta,\Omega_R)-\Omega_R^2\cdot\Omega\cdot\left(\Delta-\Omega-i\displaystyle{\frac{\gamma}{2}}\right)\right]%
\nonumber\\%
\nonumber\\%
&&+\gamma\bar{\sigma}_{+}\cdot\Omega_R\left(\Delta+\Omega+i\displaystyle{\frac{\gamma}{2}}\right)%
\left[\left(\Delta-\Omega-i\displaystyle{\frac{\gamma}{2}}\right)M^{\ast}(\Omega;\gamma,\Delta,\Omega_R)-\Omega_R^2\cdot\Omega\cdot\left(\Delta-\Omega+i\displaystyle{\frac{\gamma}{2}}\right)\right]%
\label{b.11}%
\end{eqnarray}
and
\begin{eqnarray}
N^{(++)}(\Omega)&=&\frac{\gamma}{2}\,\Omega_R^2\left(\Delta-\Omega + i\displaystyle{\frac{\gamma}{2}}\right)M(\Omega;\gamma,\Delta,\Omega_R)%
+\frac{\gamma}{2}\,\Omega_R^2\left(\Delta+\Omega+i\displaystyle{\frac{\gamma}{2}}\right)M^{\ast}(\Omega;\gamma,\Delta,\Omega_R)%
-\gamma\cdot\Omega_R^4\cdot\Delta\cdot\Omega%
\nonumber\\%
\nonumber\\%
&&+\gamma\bar{\sigma}_{-}\cdot\Omega_R\left(\Delta+\Omega-i\displaystyle{\frac{\gamma}{2}}\right)\left(\Delta-\Omega + i\displaystyle{\frac{\gamma}{2}}\right)%
M(\Omega;\gamma,\Delta,\Omega_R)%
+\gamma\bar{\sigma}_{-}\cdot\Omega_R\left(\Delta-\Omega-i\displaystyle{\frac{\gamma}{2}}\right)\left(\Delta+\Omega + i\displaystyle{\frac{\gamma}{2}}\right)%
M^{\ast}(\Omega;\gamma,\Delta,\Omega_R)%
\nonumber\\%
\nonumber\\%
&&-\gamma\bar{\sigma}_{-}\cdot\Omega_R^3\cdot\Omega\left(\Delta+\Omega-i\displaystyle{\frac{\gamma}{2}}\right)\left(\Delta-\Omega-i\displaystyle{\frac{\gamma}{2}}\right)%
-\gamma\bar{\sigma}_{+}\cdot\Omega_R^3\cdot\Omega\left(\Delta+\Omega+i\displaystyle{\frac{\gamma}{2}}\right)\left(\Delta-\Omega+i\displaystyle{\frac{\gamma}{2}}\right)%
\label{b.12}%
\end{eqnarray}
\end{widetext}
The denominator and numerator under the integrand of the transverse component (\ref{b.7}) are, respectively, given by
\begin{equation}
D_{\perp}(\Omega)=\left|4\left(\Delta+\Omega+\displaystyle{i\frac{\gamma}{2}}\right)(\Omega+i\gamma)-\Omega_R^2\right|^2%
\label{b.13}%
\end{equation}
and
\begin{eqnarray}
N_{\perp}(\Omega)&=&16\gamma\,\left(\Omega^2+\gamma^2\right)+4\Omega_R^2\gamma\,\left(\frac{1}{2}+\langle\sigma_Z\rangle\right)%
\nonumber\\%
&+& 8\Omega_R\gamma\,(\Omega+i\gamma)\langle\sigma_{-}\rangle + 8\Omega_R\gamma\,(\Omega-i\gamma)\langle\sigma_{+}\rangle%
\nonumber\\%
\label{b.14}%
\end{eqnarray}
Although this turns us to quite cumbersome representation of the susceptibility tensor, it can be structured and understood as a superposition of the separated poles after evaluation of the Cauchy integrals. 

\subsection{General analysis and asymptotic behavior}

\noindent The pole positions are determined by the denominator roots of (\ref{b.8}), (\ref{b.9}) and (\ref{b.13}) appearing as polynomial functions of $\Omega$. These roots specify the position of original atomic resonances as well as clarify the quasi-energy structure associated with the Mollow triplet. For a $z$-polarized probe the latter are given by the roots of the following cubic equations
\begin{equation}
M(\Omega,\gamma,\Delta,\Omega_R)\equiv -(\Omega-\Lambda_1)(\Omega-\Lambda_2)(\Omega-\Lambda_3)=0%
\label{b.15}
\end{equation}
where $\Lambda_{m}=\Omega_m-i\Gamma_{m}$, with $m=1,2,3$ exactly coincide with the resonances of nonlinear fluorescence emitted by the atom under action of the control field only. However, in responding on $x$ and $y$ polarized probe the same resonance features have different locations because the upper $|x\rangle$ and $|y\rangle$ states are not distorted by the entire interaction process, see Fig.~\ref{fig1}. Then the roots of (\ref{b.14}) differ from the roots of (\ref{b.8}) and (\ref{b.9}).

\begin{figure}[t]
\includegraphics[width=8.5cm]{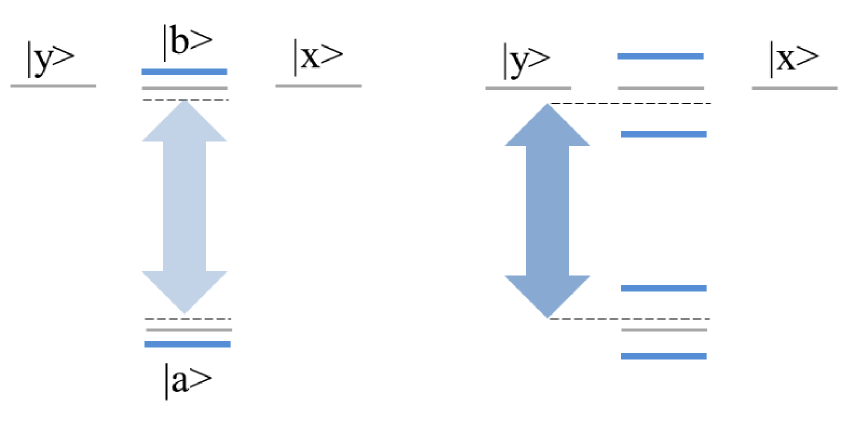}%
\caption{Left: The energy structure distorted by action of a relatively weak and red detuned control field. Atom preserves its original spectrum (shown by gray bars in both the diagrams), but the energies of states $|a\rangle$ and $|b\rangle$ are slightly shifted and broadened. Right: The action of the strong control field, saturating the transition $|a\rangle\to|b\rangle$, transforms its spectrum to the quasi-energy structure of Mollow-triplet.}
\label{fig6}%
\end{figure}

The situation becomes even more subtle, since the resonances can interfere with each other and may have only imaginary parts in some cases. Indeed, the triplet structure appears only after overcoming the threshold condition $\Omega_R>\gamma/4$, otherwise all $\Lambda_{m}$ have only imaginary parts. Furthermore, the central resonance peak always is imaginary valued such that $\Omega_2=0$. Hence in the case of $\Delta=0$ the central resonance overlaps the frequency of the undisturbed atomic transition. The transformation of the atomic energy structure from slightly distorted to the qusi-energy Mollow-triplet is shown in Fig.~\ref{fig6}  

Evaluation of integrals (\ref{b.5})-(\ref{b.7}) would be hard to do analytically, and, in general, it can be done by a round of numerical simulations. Nevertheless we can clarify they asymptotic behavior in the limits of weak and strong nonlinearity.

\subsubsection{Weak nonlinearity}
\noindent In the limit $s\ll 1$ we arrive at the following approximations of integrals (\ref{b.5})-(\ref{b.7}) 
\begin{eqnarray}
\lefteqn{\chi_{zz}^{(+-)}(\Omega)}%
\nonumber\\%
&&\approx-\frac{n_0d_0^2}{\hbar}\;\frac{1}{\Omega+\Delta+i\displaystyle{\frac{\gamma}{2}}+\displaystyle{\frac{\Omega_R^2}{2\left(\Delta-i\displaystyle{\frac{\gamma}{2}}\right)}}}
\left[1-\frac{\Omega_R^2}{2\left(\Delta^2+\displaystyle{\frac{\gamma^2}{4}}\right)}\right]%
\nonumber\\%
\label{b.16}%
\end{eqnarray}
and
\begin{equation}
\chi_{zz}^{(++)}(\Omega)\approx-\frac{n_0d_0^2}{2\hbar}\frac{\Omega_R^2}{\left[\left(\Omega-i\displaystyle{\frac{\gamma}{2}}\right)^2-\Delta^2\right]\left(\Delta+i\displaystyle{\frac{\gamma}{2}}\right)}%
\label{b.17}
\end{equation}
and
\begin{eqnarray}
\lefteqn{\chi_{xx}^{(+)}(\Omega)=\chi_{yy}^{(+)}(\Omega)}%
\nonumber\\%
&&\approx-\frac{n_0d_0^2}{\hbar}\;\frac{1}{\Omega+\Delta+i\displaystyle{\frac{\gamma}{2}}+\displaystyle{\frac{\Omega_R^2}{4\left(\Delta-i\displaystyle{\frac{\gamma}{2}}\right)}}}
\left[1-\frac{\Omega_R^2}{4\left(\Delta^2+\displaystyle{\frac{\gamma^2}{4}}\right)}\right]%
\nonumber\\%
\label{b.18}%
\end{eqnarray}
Expressions (\ref{b.16}) and (\ref{b.18}) contain the following two corrections to linear response. Firstly, the deviation from unity in square brackets is the result of a partial depopulation of the state $|a\rangle$ with repopulation to the state $|b\rangle$, induced by the control field. This process can be associated with Kerr-type nonlinearity of the medium. Secondly, the interaction with the control field has a dressing effect and shifts and broadens both the energy levels, as visualized in the left diagram of Fig.~\ref{fig6}.

The susceptibility component (\ref{b.17}) shows the parametric conversion and enhancement, taking place between the two phase conjugated modes, the signal-probe ($\omega_c+\Omega$) and idler-probe ($\omega_c-\Omega$), which in the regime of weak nonlinearity is linearly dependent on a power of the control field, playing the role of pump field in the four-wave mixing process. Both modes contribute equally to this process, so the detuning $\Omega$ can be as positive as negative. In this regime, the medium can be adjusted to work as an ideal parametric amplifier.

\subsubsection{Saturation regime}
\noindent In the strong saturation limit $s\gg 1$ the spectral location of the triplet components are given by
\begin{eqnarray}
\Lambda_1&\equiv&-\Lambda^\ast\sim -\Omega_R-i\frac{3}{4}\gamma%
\nonumber\\%
\Lambda_2&\sim&-i\frac{\gamma}{2}
\nonumber\\%
\Lambda_3&\equiv&+\Lambda\sim +\Omega_R-i\frac{3}{4}\gamma%
\label{b.19}%
\end{eqnarray}
However, as clear from the right diagram in Fig.~\ref{fig6}, the Mollow-triplet structure attributes only the $z$-polarized probe. For $x$ and $y$ polarizations only the ground state splitting can be observed as an Autler-Towns resonance doublet.  

Let us first focus on the residue contributions to the integrals (\ref{b.5}), (\ref{b.6}) coming from the pole points (\ref{b.19}). Then near the central resonance feature we subsequently obtain
\begin{equation}
\chi_{zz}^{(+-)}(\Omega)\approx\frac{n_0d_0^2}{2\hbar}\;\frac{i\gamma\left(\Delta+i\displaystyle{\frac{\gamma}{2}}\right)}%
{\Omega_R^2\left(\Omega+i\displaystyle{\frac{\gamma}{2}}\right)}%
\label{b.20}%
\end{equation}
and
\begin{equation}
\chi_{zz}^{(++)}(\Omega)\approx\frac{n_0d_0^2}{2\hbar}\;\frac{i\gamma\,\left(\Delta-i\displaystyle{\frac{\gamma}{2}}\right)}%
{\Omega_R^2\left(-\Omega+i\displaystyle{\frac{\gamma}{2}}\right)}%
\label{b.21}%
\end{equation}
which indicates vanishing of both the components as $O(1/s)$ in the saturation limit.

Near the blue sideband $\omega\sim\Lambda_3=\Lambda$ we obtain
\begin{equation}
\chi_{zz}^{(+-)}(\Omega)\approx \frac{n_0d_0^2}{2\hbar}\frac{\Delta+i\displaystyle{\frac{\gamma}{2}}}{\Omega_R(\Omega-\Lambda)}%
\label{b.22}%
\end{equation}
and
\begin{equation}
\chi_{zz}^{(++)}(\Omega)\approx -\frac{n_0d_0^2}{2\hbar}\frac{\Delta-i\displaystyle{\frac{\gamma}{2}}}{\Omega_R(-\Omega-\Lambda)}%
\label{b.23}%
\end{equation}
which vanish as $O(1/\sqrt{s})$ in the saturation limit. The asymptotic behavior near the red sideband can be found from (\ref{b.22}) and (\ref{b.23}) by replacement $\Lambda\to -\Lambda^{\ast}$. 

For the response in the $x$ and $y$ polarizations, we arrive at the following 
\begin{widetext}
\begin{equation}
\chi_{xx}^{(+)}(\Omega)=\chi_{yy}^{(+)}(\Omega)\approx-\frac{n_0d_0^2}{4\hbar}\left[\frac{1}{\Omega-\displaystyle{\frac{\Omega_R}{2}+\frac{\Delta}{2}+i\frac{3\gamma}{4}}}%
+\frac{1}{\Omega+\displaystyle{\frac{\Omega_R}{2}+\frac{\Delta}{2}+i\frac{3\gamma}{4}}}\right]%
\label{b.29}%
\end{equation}
\end{widetext}
which tells us that any probe, in transverse polarization, scatters from the atom populating the state $|a\rangle$ with one-half probability and equally weighted by the doublet components.The transverse susceptibility (\ref{b.29}) does not vanish in the saturation limit, but becomes off-resonant with (\ref{b.21}) and (\ref{b.22}). This is clearly foreseen from the right diagram in Fig.~\ref{fig3} and is a certain manifestation of the Autler–Townes effect observed by probing the system on an adjacent transition.  

The main consequence, following from the above derivation, is that the parametric process has an optimal regime for some specific value of the saturation parameter $s$, This optimum can be clarified by numerical calculations, and the respective results are presented in the main text. In the saturation limit the most effective coupling of the two phase conjugated modes takes place near the Mollow sidebands.

\subsection*{Data Availability}
\noindent Data available on request from the authors.

\nocite{*}
\bibliography{reference}

\end{document}